\documentclass[english,prb,floats,superscriptaddress,usenames]{revtex4}
\usepackage[T1]{fontenc}
\usepackage[latin9]{inputenc}
\usepackage{amsmath}
\usepackage{amssymb}
\usepackage{graphicx}
\usepackage{esint}
\usepackage{lipsum}

\makeatletter

\newcommand{\be}{\begin{equation}} 
\newcommand{\ee}{\end{equation}}
\newcommand{\bea}{\begin{eqnarray}}   
\newcommand{\eea}{\end{eqnarray}}
\newcommand{\ep}{\epsilon}

\usepackage{babel}

\makeatother

\usepackage{babel}
\begin{document}

\date{\today}


\title{Heat, temperature and Clausius inequality in a model for active Brownian particles }

\author{Umberto Marini Bettolo Marconi}
\address{ Scuola di Scienze e Tecnologie, 
Universit\`a di Camerino, Via Madonna delle Carceri, 62032, Camerino, INFN Perugia, Italy}
\author{Andrea Puglisi}
\address{Consiglio Nazionale delle Ricerche-ISC, Rome, Italy}

\author{Claudio Maggi}

\address{NANOTEC-CNR, Institute of Nanotechnology, Soft and Living Matter Laboratory, Piazzale A. Moro 2, I-00185, Roma, Italy }


\date{\today}							

\begin{abstract}
Methods of stochastic thermodynamics and hydrodynamics are applied to
the a recently introduced model of active particles. The model consists of an overdamped
particle subject to Gaussian coloured noise. Inspired by stochastic thermodynamics, we
derive from the system's Fokker-Planck equation the average exchanges
of heat and work with the active bath and the associated entropy
production. We show that a Clausius inequality holds, with the local
(non-uniform) temperature of the active bath replacing the uniform
temperature usually encountered in equilibrium systems.  Furthermore,
by restricting the dynamical space to the first velocity moments of
the local distribution function we derive a hydrodynamic description
where local pressure, kinetic temperature and internal heat fluxes
appear and are consistent with the previous thermodynamic analysis.
The procedure also shows under which conditions one obtains the
unified coloured noise approximation (UCNA): such an approximation
neglects the fast relaxation to the active bath and therefore yields
detailed balance and zero entropy production.  In the last part,  by using multiple
time-scale analysis, we provide a
constructive method (alternative to  UCNA) to determine the solution of the Kramers equation
and go beyond the detailed balance condition determining negative entropy production.

\end{abstract}
\maketitle

\section*{Introduction}

Recently, there has been an upsurge of interest in active matter
systems made of self-propelled particles which take energy from the
environment to sustain their motion
\cite{ramaswamy2010mechanics,marchetti2013hydrodynamics}. There are several reasons why this
subject has drawn the attention of biologists and physicists: active
particles model not only living systems such as Escherichia coli
bacteria, spermatozoa, swarms of animals etc., but also manmade
inanimate objects such as Janus spherical particles with catalytic
patches coatings, polymeric spheres encapsulating a hematite cube,
rod-shaped particles consisting of Pt and Au segments
\cite{bechinger2016active} which can be studied in a laboratory.
Since the constituents of active matter are powered by some external
engine and constantly spend energy to move through a viscous medium,
they are permanently out of equilibrium and thus provide new
challenges in non equilibrium statistical mechanics~\cite{S16}.  Every element of
an active matter system can be considered out of equilibrium, in
contrast to boundary driven systems, like a system subject to a
concentration gradient which is locally equilibrated.  

In contrast with passive Brownian particles subject to thermal
fluctuations, such as colloids in solution, active systems can be
described as assemblies of particles driven by fluctuating forces
which are generically correlated in time. Theoreticians have proposed
several descriptions of the active dynamics including the
run-and-tumble model \cite{berg2004coli,tailleur2008statistical}, the active
Brownian particle model \cite{romanczuk2012active,stenhammar2014phase,cates2013active}  and the Gaussian colored noise (GCN) model \cite{fily2012athermal,farage2015effective},
where the direction of motion fluctuates, but on a short-time scale there is a
persistence to move in the current directions~\cite{peruani2007self}.  The persistent character of their
trajectories is measured experimentally through their diffusivity
\cite{bechinger2016active}, which is usually much larger than the diffusivity of
colloidal particles.  The last model is a particularly simple
description of the self-propulsion mechanism, obtained by considering
the motion of the particles as a set of coupled Langevin
equations subject to a noise with a correlation time $\tau>0$,
replacing the white noise ($\tau=0$) characterising passive matter.
Thanks to this simple mathematical structure one can obtain many
results in an explicit form.  In particular, in the present paper we
perform a study of the non equilibrium stochastic energetics of the
GCN model and extend the analysis beyond the steady state and zero
current regime which has been the subject of previous work of some of
the authors~\cite{marconi2015towards,marconi2015velocity}.

Stochastic energetics aims to provide a link between stochastic
processes - which constitute an effective dynamical description of
mesoscopic systems (i.e. where the degrees of freedom of thermostats
are replaced by stochastic effective terms) - and
thermodynamics~\cite{sekimoto2010stochastic,BPRV08}.  It basically
answers the questions of what are the heat and the work associated
with a random motion and aims at tracing the role of the energy
exchanges in the variation of the Gibbs entropy~\cite{LS99,seifert05}.
In macroscopic thermodynamics, one studies the change of thermodynamic
entropy $S$ after a transformation from the equilibrium state $A$ to
equilibrium state $B$, during which a heat $\int_{A \to B} \delta Q$
goes from the thermostat to the system. For such a transformation one
has
\begin{equation}
S(B)-S(A)=\int_{A \to B} \frac{\delta Q}{T} + \Sigma(A \to B)
\end{equation} 
where $\Sigma$ - known as entropy production - satisfies $\Sigma \ge 0$ with the equal
sign valid only in a quasi-static transformation. In a cyclic
transformation $S(B)=S(A)$, so that the Clausius inequality is recovered, $\oint \delta Q/T=-\Sigma \le 0$.

In stochastic thermodynamics the equilibrium entropy is replaced by
Gibbs entropy $s(t)$ (details are given hereafter) and the average instantaneous entropy
variation is written as
\begin{equation}
\frac{ds}{dt} = \dot{s}_m + \dot{s}_s,
\end{equation}
with the entropy production rate of the system $\dot{s}_s \ge 0$ ($=0$
only at equilibrium) and $\dot{s}_m$ interpreted as ``entropy 
production of the surrounding medium''~\cite{seifert05}. Indeed in
many cases of non-conservative forces applied to a system in contact
with a heat bath at temperature $T$, it appears that
$\dot{s}_m=\dot{q}/T$, with $\dot{q}$ the average heat flux going from
the heat bath to the system~\footnote{In the rest of the paper we use,
  for simplicity, the notation $\dot{q}$ to indicate a
  heat flux, which of course does not imply the existence of an observable $q$
  depending on the state of the system.}. In the stationary state, therefore, one has - again -
\begin{equation} \label{clausius}
\frac{\dot{q}}{T}=-\dot{s}_s \le 0.
\end{equation}

There are however other cases where such a simple interpretation of
$\dot{s}_m$ is lost. One of the most studied cases is when the
non-conservative force depends on the velocity of the particle, a fact
which is common in mesoscopic systems with
feedback~\cite{kim2004entropy} and in some models of active particle~\cite{sarracino,ganguly2013stochastic,chaudhuri2014active,FNCTVW16}.
In those examples one finds a more complicate structure of the kind
\begin{equation}
\frac{ds}{dt}=\frac{\dot{q}}{T}+\dot{s}_s+\dot{s}_{act}\\
\end{equation}
where still $\dot{s}_s \ge 0$, but an additional contribution
$\dot{s}_{act}$   entropy production  appears without
a well defined sign \footnote{ We use ``act'' for ``active'' , but previous
authors have called it ``pumping''  \cite{celani2012anomalous}} . As a consequence in the stationary state the
Clausius inequality is no more guaranteed. A possible interpretation of this fact
is a problem in the modellisation of the external non-conservative
agent~\cite{CP15}.

Here we apply the methods of stochastic energetics and thermodynamics
to the GCN model, showing that the instantaneous entropy variation can
be written as
\begin{subequations} \label{mainres}
\begin{align} 
\frac{ds}{dt}&=\int dx \frac{\dot{\tilde{q}}(x)}{\theta(x)}+\dot{s}_s \\
\dot{q}&=\int dx \dot{\tilde{q}}(x),
\end{align}
\end{subequations}
with $\dot{s}_s \ge 0$ taking a simple Onsager-like structure and
$\dot{\tilde{q}}(x)$ representing the local energy flux coming from the 
active bath (represented by the Gaussian coloured
noise) which has an effective local temperature
$\theta(x)$ (details in Section Model and Methods). Remarkably, Eq.~\eqref{mainres} generalises the Clausius inequality~\cite{BGJL13,MN14,BDGJL15}, as in the stationary state it implies
\begin{equation} \label{newcla}
\int dx \frac{\dot{\tilde{q}}(x)}{\theta(x)} \le 0.
\end{equation}

Interestingly, a known approximation of the GCN model where only the
longer time-scales are considered (``UCNA'' approximation \cite{hanggi1995colored,jung1987dynamical,h1989colored}), yields a zero entropy production,
being $\dot{q}(x) = 0$ everywhere, as in local
equilibrium. This is consistent with a known observation: coarse-graining operations that remove fast
time-scales are likely to suppress part - or all
- of the entropy production of a system~\cite{PPRV10,CPV12,esposito2012stochastic,celani2012anomalous,bo2014entropy}. For the same reason,
the entropy production needed to sustain the active bath (see for
instance~\cite{S16}), which is usually {\em hotter} then the equilibrium (solvent) bath, does not appear in our description here,
because the model is defined on a time-scale which is slower than the
one of the molecular heat exchanges.

Eqs.~\eqref{mainres}-\eqref{newcla} are not the only important result of our work. We
also derive the hydrodynamic equations of the GCN model (for
non-interacting particles or, equivalently, in the dilute limit) for
density, momentum, and temperature local fields. This is pivotal in getting
further insight in the local thermodynamics of the model, i.e. making
explicit terms such as the local pressure, local kinetic temperature,
internal heat fluxes and local entropy production. 
 
This paper is organised as follows: in section Model and Methods, we present
the model of independent active particles in one dimension subject to
a generic potential and write the evolution equation for the Kramers
equation for the position-velocity distribution function.  Using the Kramers equation and after introducing the
concepts of work, heat and Gibbs entropy, we present a thermodynamic
description of the non equilibrium coloured noise system by relating
the entropy variation to the power and heat flux of the system.  Using
stochastic thermodynamic methods we show that there exists a local
Clausius inequality involving the
heat flux and a non uniform local temperature.  Such a scenario is
corroborated by the analysis of the  hydrodynamic equations
derived from the Kramers equation.  We also derive the UCNA approximation
 from the hydrodynamic equations by taking the overdamped limit and show that
this approximation corresponds to a vanishing entropy production which
is consistent with the fact that the UCNA satisfies the detailed balance condition. 
To go beyond such an approximation
we use a multiple-time scale method and show
how a finite entropy production arises.   Finally, we 
present our conclusions.

The appendix  illustrates the   multiple-time scale method  used in the main text to 
expand   the Kramers equation in powers of $\sqrt{\tau}$
around the $\tau=0$ solution and derive the form of the phase-space probability
distribution without imposing the detailed balance condition.

\section*{Model and methods}
\label{sec1}

We consider a minimal model describing the basic dynamical properties of a
suspension of mutually non-interacting active particles in the
presence of an external field.  The steady state properties of such a
model, including interactions, have been recently studied in a series of papers, but little
attention has been devoted to its dynamical properties.  For the sake
of simplicity we study a one dimensional model, and leave the
straightforward extension to higher dimensions with interactions to future work.  
 As mentioned in the introduction,  the distinguishing feature of active matter is the ability of the self-propelled particles  to convert energy from the environment into persistent motion so that one observes a variety of peculiar properties such as extraordinary large diffusivities as compared to suspensions of colloidal particles of similar size and spatial distributions not following Boltzmann statistics, to mention just a few.  Besides the active Brownian particle (ABP) model, which considers 
independently the translational and the rotational degrees of freedom of the particles, the GCN has recently gained 
increasing popularity among theoreticians, because it lends itself to more analytical treatments. 
Farage et al. \cite{farage2015effective}  presented a clear discussion of how the GCN can be considered as a coarse grained version of the ABP, and can be derived by averaging over the angular degrees of freedom.
The essence of the GCN is to describe the dynamics of an active
suspension by means of an over damped equation for the position variable subject to a deterministic force and to a stochastic
driving. At variance with colloidal solutions described by standard Brownian dynamics  the noise is time-correlated to account for the persistence of the trajectories. 
We model the
effective dynamics for the space coordinates of an assembly of non-interacting active
Brownian particles \cite{maggi2015multidimensional,marconi2015towards}
as 
\begin{equation}
\label{uno}
\dot{x}(t) =\frac{1}{\gamma}f(x) + a(t) 
\end{equation}
where the term $a$, also called ``active bath'', evolves according to the law:
\begin{equation}
\dot a(t) =- \frac{1}{\tau} a(t) + \frac{ D^{1/2}}{ \tau} \eta (t)  .
\label{due}
\end{equation}
The force $f(x)$ acting on each particle is  {\bf time-independent} and associated to
the potential $w(x)$, $\gamma$ is the drag coefficient, whereas
the stochastic force $\eta(t)$ is a Gaussian and Markovian process distributed with zero mean
and moments  $\langle \eta(t) \eta(t')\rangle=2    \delta(t-t')$.  
The coefficient $D$ due to the activity is related to the correlation of the 
Ornstein-Uhlenbeck process  $a(t)$ via 
$$\langle a(t) a(t')\rangle=\frac{D}{\tau} \exp( -\frac{|t-t'|}{\tau}) .$$ 
In order to proceed analytically it is convenient to switch from the $(x,a)$ variables to the phase-space variables $(x,v)$ 
where $v=\dot x$ and rewrite \eqref{uno} and \eqref{due} as
\bea
\dot x&=&v
 \nonumber\\
 \dot v&=&-\frac{1}{\tau}\left(1-\frac{\tau}{\gamma} \frac{df}{dx}\right) v+ \frac{1}{\tau\gamma} f +\frac{ D^{1/2}}{ \tau} \eta
\label{sistema2}
 \eea
 One can immediately write the associated  Kramers equation  for the phase-space distribution $p(x, v;t) $:
 \be
\frac{ \partial p }{\partial t}+ v \frac{\partial p}{\partial x}+\frac{f(x)}{\gamma \tau} \frac{\partial p}{\partial v}
= \frac{1}{ \tau}  \frac{\partial }{\partial v} \Bigl(  \frac{D}{\tau}\frac{\partial }{\partial v}+    \Gamma(x) v \Bigl) p
\label{kunob}
\ee
with $\Gamma(x)=(1-\frac{\tau}{\gamma} \frac{ d f}{dx})$. 
The second and third term on the  left hand side represent the streaming terms in the evolution,
that is correspond to the Hamiltonian evolution of the phase-space distribution, whereas the right hand side
describes the dissipative part of the evolution.  
Notice that, at variance with the standard  Kramers equation the force is divided by the unusual factor $\tau\gamma$
and the friction is space dependent and  varies with $\tau$.

\subsubsection*{Transport equation in phase-space and entropy production}
\label{sec2}

In order to proceed further, it is time-saving to adopt non-dimensional variables for positions, velocities,  and time
and rescale forces accordingly.
We define $v_T=\sqrt{D/\tau}$ and  introduce 
the following non-dimensional variables: 
\begin{equation}
\bar t\equiv t\frac{v_T}{l}, \qquad V\equiv\frac{v}{v_T}, 
\qquad X\equiv\frac{x}{l},\qquad   F(X)\equiv\frac{l f(x)}{D\gamma}    ,\qquad \zeta=\frac{l}{\tau v_T}
\label{adim1}
\end{equation}
Interestingly, $\zeta$ is the inverse of the
P\'eclet number, $Pe=\sqrt{D\tau}/l$,
of the problem,  that is the ratio between  the mean square diffusive displacement due to the active bath in a time interval $\tau$,
the so called persistence length, over the typical length of the problem, $l$, such as length-scale of the variation of the external potential $w(x)$.
As it will be clear in the following the parameter $\zeta$ plays the role of a non-dimensional friction.

We rewrite  Kramers' evolution equation for 
the phase-space distribution function
using  (\ref{adim1}) 
as:
\begin{equation}
\frac{\partial  P(X,V,\bar t)}{\partial \bar t} +V \frac{\partial }{\partial X}  P(X,V,\bar t) 
+ F(X) \frac{\partial }{\partial V}  P(X,V,\bar t)
=\zeta \frac{\partial}{\partial V}\Bigl[
\frac{\partial }{\partial V }+g(X) V\Bigr]   P(X,V,\bar t) \, ,
\label{kramers0b}
\end{equation}
where $
g(X)=\Gamma(x) =1-\frac{1}{\zeta^2}\frac{d F}{dX}
$.

  
  \subsubsection*{Work, heat and entropy production}
  Equation \eqref{kramers0b} can be written as a continuity equation in phase space
  \be
   \frac{\partial}{\partial \bar t} P(X,V,\bar t)  +  \frac{\partial }{\partial X} I_x(X,V,\bar t) + \frac{\partial }{\partial V}  I_v (X,V,\bar t)   =0
   \label{continuitya}
  \ee
  by introducing a probability current vector,  $(I_x,I_v)$, whose components are  the sum of a reversible current  
  (indicated with a superscript R)
  \be
  (I^R_x,I^R_v)\equiv(VP,  F  P )
  \label{reversiblecurrent}
  \ee
  and a dissipative or irreversible current   (indicated with a superscript D)
  \be
  (I^D_x,I^D_v)\equiv(0,  -\zeta  \frac{\partial P}{\partial V}-\zeta g(X) V P )
  \label{irreversiblecurrent}
  \ee

 The dynamical equation \eqref{kramers0b} ruling the statistical
 evolution of the phase space distribution can be seen as the result
 of an Hamiltonian (non dissipative) dynamics coupled to an heat-bath:
 \be \frac{\partial}{\partial \bar t} P + \frac{\partial E}{\partial
   V} \frac{\partial }{\partial X} P -\frac{\partial E}{\partial X}
 \frac{\partial }{\partial V} P=\zeta  \frac{\partial}{\partial V}\Bigl[
\frac{\partial }{\partial V }+g V\Bigr]  P \ee where \be E(\bar
 t)=\frac{1}{2} V^2 +U(X) \ee with $w= D\gamma U$ such that
 $F(X)=-d U/dX$. 
 Upon differentiating the
 expectation value of $E(t)$ with respect to time we obtain 
\be
 \langle \dot E(\bar t)\rangle =\langle \dot W( \bar t)\rangle
 +\langle \dot Q( \bar t)\rangle 
\ee 
with 
\bea \langle \dot W(\bar t)\rangle &=& \iint dX dV P(X,V,\bar t) \frac{\partial}{\partial \bar t} U(X)=0 \\ 
\langle \dot Q( \bar t)\rangle &=& \iint dX dV  E(\bar t) \frac{\partial}{\partial \bar t} P(X,V,\bar t) \\ 
&=& -\iint dX dV E(\bar t) \Bigl [\frac{\partial }{\partial X}  I_x(X,V,\bar t) + \frac{\partial }{\partial V} I_v (X,V,\bar t) \Bigr] \\ 
&=& \iint dX dV \Bigl [\frac{\partial E}{\partial X}    I_x(X,V,\bar t) + \frac{\partial E }{\partial V} I_v (X,V,\bar t)   \Bigr] 
\eea 
Since in the simple case of a time independent potential (which is the
situation considered in the rest of this paper), $\langle \dot{W}
\rangle =0$, i.e. $\langle \dot E \rangle = \langle \dot{Q} \rangle$.
Explicitly, we obtain 
\be 
\langle \dot Q( \bar t)\rangle =-\zeta \iint dX dV \left[g(X) V^2 P+V \frac{\partial P   }{\partial V} \right] = \iint dX dV V I^D_v .
\label{qpunto}
\ee 
 We underline that our definition of heat is
  coherent with the standard definition of stochastic
  energetics~\cite{sekimoto2010stochastic}. Indeed it is
  straightforward to verify that $\langle \dot{Q} \rangle$ corresponds
  to the average of the power dissipated by forces acting on the
  particle which are related to the active bath, i.e. all forces
  appearing in Eq.~\eqref{sistema2} but the conservative term
  $f/\tau\gamma$.

We introduce now the Gibbs entropy $S(\bar    t)\equiv -\iint dX dV
P(X,V,\bar t) \ln P(X,V,\bar t) $, which in equilibrium systems
connects the statistical level and the  probability distribution
$P(X,V,\bar t)$ to the macroscopic thermodynamic quantities such as
heat and work, and relate its time derivative to the heat rate
$\langle \dot Q(\bar t)\rangle $.
We consider:
  \be 
 \dot  S(\bar t)=   -\iint dX dV  \frac{\partial}{\partial \bar t} P(X,V,\bar t)  \ln P(X,V,\bar t)  
 =   \iint dX dV \,{\bf \nabla}\cdot    \Bigl[{\bf I}^R+{\bf I}^D \Bigr]    \ln P(X,V,\bar t)
\label{gibbsentropy}
\ee
 To derive equation  \eqref{gibbsentropy}  we used the continuity equation \eqref{continuitya} .
  After integrating by parts and using the zero flux boundary conditions at infinity and  the definitions \eqref{reversiblecurrent} and  \eqref{irreversiblecurrent}  we obtain the expression:
\begin{align}
\dot S(\bar t)  &=-  \iint dX dV \,  \frac{1}{P} \nabla P    \cdot [{\bf I}^R+{\bf I}^D] = \dot{S}^R + \dot{S}^D \\
\dot{S}^{R(D)} &= \iint dX dV \,  P   \nabla \cdot \frac{[{\bf I}^{R(D)}]}{P}.
\end{align}
 One can see that the reversible contribution to $\dot S$ vanishes since
  \be 
 \dot S^R(\bar t)  =   \iint dX dV \,  P  (X,V,\bar t)   \nabla \cdot \frac{{\bf I}^R (X,V,\bar t)  }{P (X,V,\bar t) }            =-  \iint dX dV \, P \Bigl[  \frac{\partial}{\partial X}  \frac{\partial  E(X,V)}{\partial V}   -  \frac{\partial}{\partial V} \frac{\partial E(X,V)}{\partial X} \Bigr]=0.
  \ee
 Analogously we may write the dissipative  contribution to $\dot S$ as
 \be 
 \dot S^D(\bar t)= \frac{1}{\zeta}  \iint dX dV \left[\frac{ ( I_v^D)^2}{P}+  \zeta g(X) V I^D_v \right]
  \ee
so that the total time derivative of the entropy \eqref{gibbsentropy} turns out to be:
\begin{subequations}  \label{equazionesekimoto}
\begin{align}
\dot S(\bar t) &= \dot{S}_{s}(\bar t) + \dot{S}_{m}(\bar t) \\
\dot{S}_s(\bar t) &= \zeta \iint dX dV \frac{1}{P} \Bigl(  \frac{\partial P}{\partial V}+g(X) VP   \Bigr)^2 \\
\dot{S}_m(\bar t) &= \iint dX dV g(X) V I^D_v = -\zeta \iint dX dV g(X) \Bigl( g(X)V^2 P +V  \frac{\partial P}{\partial V}  \Bigr) .
\label{essem}
\end{align}
\end{subequations}
It is clear that in the steady state the time derivative of the
entropy vanishes $\dot S=0$,  as well as  $\langle \dot{Q} \rangle$ . However, it is interesting to identify $\dot{S}_s$ as an
entropy production rate, which is indeed always non-negative, and $\dot{S}_m$ as the flux of entropy due to heat exchanges between
the system and the surroundings, also known as entropy production of the medium~\cite{seifert05}.
 One immediately notices that $\dot{S}_m$ is not simply proportional to  $\langle \dot Q(\bar t)\rangle$, see Eq. \eqref{qpunto}, as one would find 
 when studying the entropy production of a system coupled to an equilibrium thermostat and driven out-of-equilibrium by non-conservative forces~\cite{seifert05}.
 Here we have a spatial  distribution of temperatures. In order to appreciate that, we need to discuss the role of $g(X)$ as an inverse temperature and to
 consider the dimensional form of the equations, which is done in the following two subsections.

\subsubsection*{Absence of detailed balance condition in the GCN}
\label{secdb}
The conditions of detailed balance can be written in terms of components of the 
 dissipative or irreversible current  \eqref{irreversiblecurrent}, which  must vanish
 in the steady state~\cite{risken}:
  \be
  \left(0,  -\zeta  \frac{\partial P_s(X,V)}{\partial V}-\zeta g(X) V P_s(X,V) \right)=(0,0),
  \label{detailedb}
  \ee
being $P_s(X,V)$ the stationary distribution. Without loss of generality, such a solution can be written as the product of a weight function $\pi(X,V)$ and a ``local'' Maxwellian with velocity variance which is position-dependent:
  \be
  P_s(X,V) =\pi(X,V) \frac{g^{1/2}(X)} {\sqrt{2 \pi}}\exp(-\frac{g(X)}{2} V^2).
  \label{gauss}
  \ee 
Formula~\eqref{gauss} - inserted in Eq.~\eqref{detailedb} - requires
\begin{equation}
P_s(X,V) \frac{\partial \ln \pi(X,V)}{\partial V}=0,
\end{equation}
which implies that $\pi(X,V)=\pi(X)$ is a function of positions only and can be considered as a local density.

Instead, the vanishing of the reversible part of the current  vector is not a necessary condition for detailed balance, 
which instead requires the vanishing of its divergence. To see that, consider the case of  an Hamiltonian system,
where detailed balance  trivially holds, but the reversible component of the current in phase space is manifestly non zero.
Indeed, in the steady state  in virtue of  the Hamilton equations the associated divergence vanishes, thus confirming
our statement.
In the present case, we write the divergence of the reversible current (see Eq.~\eqref{reversiblecurrent}) as:
  \be
div {\bf I}^R= \left[ V \frac{\partial }{\partial X}  +F (X)  \frac{\partial}{\partial V} \right]  P_s(X,V) =0 .
\label{zerodiv}
\ee
Plugging the distribution \eqref{gauss} into equation \eqref{zerodiv}  and taking into account 
that the form of $P_s(X,V)$ implies the relation $ \frac{1}{g(X)} = \langle V^2\rangle_X$, where the last average
indicates the mean square value of the velocity evaluated at position $X$,
we can see that a density distribution $\pi(X)$ satisfying  \eqref{zerodiv}  does not exist for arbitrary choices of $g(X)$. In fact,
we find:
\be
   \Bigl(     \frac{1}{\pi(X)} \frac{d \pi(X)}{dX }  -F(X) g(X)  
   +\frac{1}{2}  \frac{d g(X)}{dX }[ \langle V^2\rangle_X -V^2  ]     \Bigr)  V \pi(X) \exp(\frac{-g(X)}{2} V^2) \neq 0.
  \label{nodetail1b}
\ee
 The only case where the identity holds is $g(X)=g_0$, a  constant, which occurs for $\zeta \to \infty$
which is the equilibrium limit of the model or in the cases a) of linear or b)  parabolic potentials  $U(X)$ ), and sets the equilibrium Boltzmann solution  $\pi(X)\propto e^{-g_0 U(X)}$
for any value of $V$.
In conclusion, apart from the special case $g(X)=g_0$, the Kramers equation~\eqref{kramers0b}
 does not satisfy the detailed
balance condition. 
On the other hand,  one may verify that the first three projections in velocity space of the zero divergence condition Eq.~\eqref{nodetail1b} 
(i.e. obtained by multiplying such an equation by $(1,V, V^2)$ , respectively and integrating w.r.t. $V$) can indeed be fulfilled,
and consequently, the detailed balance condition is satisfied in this velocity subspace. 
In the case of the multiplication by the  even powers of $V$ and integration by $V$ the balance equation is trivially satisfied.
 Multiplication by $V$ and integration, instead, leads to:
 \be
   \frac{1}{\pi(X)} \frac{d \pi(X)}{dX } -  \frac{1}{g(X)}   \frac{d g(X)}{dX }-F(X) g(X)   =0
  \label{nodetail2}
\ee
  whose solution is 
  \be \label{ucna1}
  \pi(X)= {\cal N}  \exp \left( -U(X)-\frac{1}{2\zeta^2} \left(\frac{d U}{dX}\right)^2 \right) g(X).
  \ee
    Interestingly,
  as we shall illustrate in below,
 the spatial distribution in Eq.~\eqref{ucna1}  coincides with the static solution of the UCNA equation, which  is known to satisfy the detailed balance condition,  but in general,  such a condition holds only approximately (i.e. only to second order in the parameter $\zeta^{-2}$)
  for the non equilibrium steady state under scrutiny.


\subsubsection*{  Dimensional form of equations}

The dimensional form of the equations is more enlightening for a thermodynamic interpretation of the entropy production of the medium, that is the
second term on the r.h.s. of Eq.~\eqref{equazionesekimoto}.  First of
all, let us notice that the local temperature appearing in the
Maxwellian Eq.~\eqref{gauss}, in dimensional form, takes the expression $\theta(x)=D/[\tau \Gamma(x)]$. 
It is also useful to define $T_b=D/\tau$, so that $\theta(x)=T_b/\Gamma(x)$.

The dimensional form of the total energy and of  the heat flux
are respectively: \be \epsilon(t)=\frac{1}{2} v^2+
\frac{w(x)}{\tau \gamma} \, , \ee 
and
\be
\dot q=- \iint dx dv \frac{\Gamma(x)}{\tau}   \Bigl[v^2  p(x,v,t) + \theta(x) v\frac{\partial}{\partial v} p(x,v,t) \Bigr ]  
\ee
and the Gibbs entropy now reads
$ s(t)\equiv -\iint dx dv p(x,v,t)  \ln p(x,v,t)  
$.
In dimensional form 
using the local temperature the total entropy time derivative $ \dot s(t)= \dot s_s(t)+ \dot s_{m}(t)$ is the sum of 
the positive quantity
 \be
  \dot s_s(t)=\frac{1}{\tau} \frac{1}{T_b}\iint dx dv \frac{1}{p} \Bigl(  \Gamma(x) v p  +T_b \frac{\partial p}{\partial v} \Bigr)^2
 \ee
and the entropy flux
 \be
 \dot s_{m}(t)= - \iint dx dv   \frac{1}{\theta(x)} \frac{\Gamma(x)}{\tau}  \Bigl [ v^2   p(x,v,t) +\theta(x) v\frac{\partial}{\partial v} p(x,v,t)\Bigr ]  .
 \label{shf}
  \ee
  It is suggestive to rewrite
  \bea
  &&
  \dot q(t)=\int dx \dot{\tilde{q}}(x,t) 
  \\&&
 \dot s_{m}(t)=\int dx \frac{1}{\theta(x)} \dot{\tilde{q}}(x,t) \label{sm}
 \eea
 with a local density of heat flux defined as
\begin{equation} \label{tildeq}
\dot{\tilde q}(x,t) =- \frac{\Gamma(x)}{\tau}\int dv \Bigl[v^2 p(x,v,t) + \theta(x) v\frac{\partial}{\partial v} p(x,v,t) \Bigr ]  =  -\frac{1}{\tau}  \frac{T_b}{\theta(x)}  n(x,t)\left[\langle v^2\rangle_x  -\theta(x)\right ],
\end{equation}
where $n(x,t)=\int dv p(x,v,t)$ and $ n(x,t)\langle v^2\rangle_x=  \int dv v^2 p(x,v,t)$ ,
where $\langle v^2\rangle_x$ is the mean squared velocity at given position.
Expression~\eqref{sm} represents an interesting connection between the
local entropy production of the medium (or entropy flux) and the local
heat flux divided by the same local temperature
$\theta(x)=T_b/\Gamma(x)$ featuring in the approximate detailed
balance solution, Eq.~\eqref{gauss}. Such a result is to be compared
with alternative expressions for entropy production of the medium
recently derived for active
systems~\cite{chaudhuri2014active,FNCTVW16}.

Remarkably, the expression~\eqref{sm} for the entropy production of
the medium and the fact that $\dot{s}_s \ge 0$, yields for the stationary
state the following generalised Clausius inequality~\cite{BGJL13}
\begin{equation} \label{gencla}
\int dx \frac{1}{\theta(x)} \dot{\tilde{q}}(x,t) \le 0.
\end{equation}

\subsubsection*{Timescales and coarse-grained levels of description}

It is interesting to discuss the characteristic timescales existing in
the model and their role in the results obtained up to here. The model
in Eq.~\eqref{uno} has its natural interpretation as a coarse-grained
version of a more refined model where the particle has a mass $m$. The
more refined model has three main timescales: 1) $\tau_m=m/\gamma$
which is the molecular kinetic relaxation timescale, that is the
timescale of the relaxation (to the statistics of the bath) of the
velocity of the particle which is achieved only when the external
forces vanish, i.e.  $a=0$ and $f=0$ (passive colloid); 2) $\tau$
which is the persistence timescale of the active force $a(t)$; 3)
$\tau_w=\Delta x/v_T$ - where $v_T=\sqrt{D/\tau}$ and $\Delta x$ is
a characteristic length-scale of the potential $w(x)$ - which is
the time needed by the particle (roughly going at speed $v_T$
which is the active bath ``thermal'' velocity) to see the variations
of the potential $w(x)$. Apart from strange choices of the
potential (e.g. $w(x)$ varying over very small length-scales), the
natural order of the three timescales is $\tau_m \ll \tau \ll
\tau_w$. Note that $\zeta=\tau_w/\tau$ (if $l=\Delta x$) and
therefore in the perturbative scheme discussed in
the Appendix the small parameter $1/\zeta$
corresponds to $\tau_w \gg \tau$.

When there is no activity ($a=0$), one has the well-known situation
of a colloid in a potential lanscape: in that case, in view of the
usual separation $\tau_m \ll \tau_w$ (now one uses the molecular velocity to define $\tau_w=\Delta
x/v_{th}$ with $v_{th}=\sqrt{T_{solv}/m}$ and $T_{solv}$ is the equilibrium temperature of the solvent), one adopts the classical overdamping approach,
i.e. looks at the motion of the particle on a timescale intermediate
between $\tau_m$ and $\tau_w$. On such a time-scale the relaxation
of the ``velocity'' to $f(x)/\gamma$ is immediate
\begin{equation}
\dot{x}=\frac{f(x)}{\gamma} + \textrm{thermal noise}.
\end{equation}
Operatively, this is equivalent to measure a `` coarse-grained velocity'' defined as $[x(t+\Delta t)-x(t)]/\Delta
t$ with a time-lag such that $\tau_m \ll \Delta t \ll
\tau_w$. This is clearly very different from the instantaneous velocity,
i.e. the one that could be measured with $\Delta t \ll \tau_m$,
usually quite difficult in real experiments.

When there is activity $a\neq 0$, the new timescale $\tau$ allows
one to operate two different levels of coarse-grain. On a timescale
which is intermediate between $\tau_m$ and $\tau$, the
coarse-grained velocity $\dot{x} \approx [x(t+\Delta t)-x(t)]/\Delta t$ with
$\tau_m \ll \Delta t \ll \tau$ satisfies
\begin{equation} \label{seclev}
\dot{x}=\frac{f(x)}{\gamma}+a(t) + \textrm{thermal noise},
\end{equation}
which becomes our initial model definition, Eq.~\eqref{uno}, when the thermal noise is neglected (which is usually safe, as it is much smaller than  $f(x)/\gamma+a(t)$).

On a much longer timescale, i.e. when the velocity is measured using
$\tau \ll \Delta t \ll \tau_w$, there is a complete overdamping,
i.e. also the active force is averaged out and only the potential
acts, but with a re-normalized viscosity $\gamma \Gamma(x)$, i.e. the
``super-coarse-grained'' velocity of the particle immediately relaxes
to
\begin{equation} \label{thirdlev}
\dot{x}=\frac{f(x)}{\gamma \Gamma(x)} + \textrm{active noise}.
\end{equation}

The entropy production of the first, most fundamental, level of
description (real velocity, measured at $\Delta t \ll \tau_m$) could
be computed by studying its complete Fokker-Planck equation: such a
small scale entropy production - however - is not particularly useful,
as it would yield an expression which includes quantities which 
can be difficult to be measured in experiments (exceptions are presented in \cite{esposito2012stochastic,celani2012anomalous,bo2014entropy} ). The meaning of
this entropy production should be simple~\cite{S16}: an external force
(activity) is keeping the particle far from reaching thermal
equilibrium (at temperature $T_{solv}$), in particular, such a force
increases the energy of the particle. There is, naturally, an energy
flux from the external force (the bacterium's engine) to the particle
and a heat flux from the particle to the bath. Our paper disregards
such a low-level entropy production and focuses on the entropy
production at the second and third levels of description. The
situation is similar to other systems where small-scale degrees of
freedom are ignored (and - consequently - their contribution to
entropy production), for instance in granular matter~\cite{PVBTW05}.

At the second level, Eq.~\eqref{seclev} or Eq.~\eqref{uno}, the only
relaxation of velocities which can be measured is that toward the local active bath
at temperature $\theta(x)=T_b/\Gamma(x)$. Its energetic counterpart is
the total heat flux (Eq. (18) or - in dimensional form - Eq. (35) of
the main text) which is the sum (space integral) of local heat fluxes,
i.e. $\dot{q}=\int dx \dot{\tilde{q}}(x)$. In the stationary state the total heat flux
is zero: there are regions where the flux goes in
one direction (e.g. the particle is hotter than $\theta(x)$)
compensated by regions where it goes in the opposite
direction. However, even with total zero heat flux, a non-zero global
entropy flux $\int dx \dot{q}(x)/\theta(x)$ may appear: this occurs
because of the non-uniformity of the temperature.   In the special case of a constant $\Gamma(x)$ , 
$\theta(x)$ is proportional to $T_b$
and the total entropy flux becomes proportional to the total heat
flux, i.e. it vanishes. This ``second-level'' entropy production misses the entropy
produced at the finer timescale (removed by the heat flux exchanged between  the
particle and the solvent heat bath), which is there even when the
$\theta(x)$ is uniform. On the contrary, in the presence of a
 non-uniform $\theta$, one gets a
non-zero (de facto negative) total entropy flux, corresponding to positive entropy production. It is remarkable that
even the incomplete description at such a second level of description
yields a thermodynamic-like description where - as in the Clausius
relation - $\int \dot{q}(x)/\theta(x) \le 0$. In other examples of
coarse-grained out-of-equilibrium systems (e.g. granular systems~\cite{PVBTW05}, but
also different models of active particles~\cite{chaudhuri2014active}, or systems with feedback~\cite{kim2004entropy})
one has that the heat flux does not rule the entropy flux and
therefore there is nothing similar to a Clausius relation~\cite{CP15}. Models with temperature gradients
showing such a relation can be found in  refs. \cite{ge2014time,bo2013entropic}.

We conclude this discussion by considering the third level of
description, where the relaxation of the velocity to the active heat
bath is also lost, Eq.~\eqref{thirdlev}. This level corresponds to the
UCNA approximation and its approximated velocity statistics are exactly
equal, everywhere, to a local Gaussian with temperature $\theta(x)$ -
even when it is non-uniform. Consistently with the theoretical discussion,
this implies local equilibrium with the active bath, or equivalently vanishing entropy production, i.e. detailed balance.




 \subsubsection*{Internal energy  and entropy density balance in dimensional form}

An interesting perspective is offered by considering the properties of
the hydrodynamic space only, instead of the properties of the full
phase-space: this is a different kind of coarse-graining, where fast
components of the full solution $p(x,v, t)$ are neglected.  
 Details of the derivation are given in appendix A, where we used the non dimensional
 variables. 
  
We define the local density field $n(x,t)=\int dv p(x,v,t)$, the local velocity field $u(x,t)= [1/n(x,t)]\int dv v p(x,v,t)$, the
local kinetic temperature field $T(x,t)= [1/n(x,t)]\int dv  p(x,v,t) (v-u(x,t))^2$, the local pressure $\pi(x,t)=n(x,t) T(x,t)$, and the
heat flux $j_q(x,t)=\int dv  p(x,v,t) (v -u(x,t))^3/2$ and
get for the following hydrodynamic-like balance equations: 
 \bea
&&
\frac{\partial n (x, t)}{\partial  t}+\frac{\partial} {\partial x} (n (x,t) u(x,t))=0
\label{brinkman0d}
\\
&&
\frac{\partial [n(x,t) u(x,t)]} {\partial t}
+ \frac{\partial} {\partial x} (n (x,t) u^2(x,t))=-   \frac{\partial \pi(x,t)} {\partial x} 
+\frac{f(x)}{\tau \gamma}n(x, t)-\frac{1}{\tau} \frac{T_b}{\theta(x)} n(x,t) u(x,t)
\label{brinkman1d}
\\
&&
\frac{\partial T(x,t)} {\partial t}+ u(x,t) \frac{\partial} {\partial x} T(x,t)  + 2 \frac{\pi(x,t)}{n(x,t)}\frac{\partial u(x,t)} {\partial x}
+2 \frac{1}{n(x,t) }
\frac{\partial} {\partial x} j_q (x,t)=-2\frac{T_b}{\tau}  \left[\frac{T(x,t)}{\theta(x)}-1\right]
\label{brinkman2d}
\eea
The term $ \pi(x,t) \frac{\partial u(x,t)} {\partial x}$ represents a compression work per unit time. 
The terms in the r.h.s of \eqref{brinkman1d} and  \eqref{brinkman2d}  balance equations make the approach to the local values of $u(x,t)$ and $T(x,t)$ fast processes,
in contrast with the slow evolution of the density.
Defining the internal energy as $\epsilon_{int}(x,t)= \frac{T(x,t)}{2}$, we immediately get 
\be
\frac{\partial \epsilon_{int}(x,t)} {\partial t}+ u(x,t) \frac{\partial} {\partial x} \epsilon_{int}(x,t)  +  \frac{\pi(x,t)}{n(x,t)}\frac{\partial u(x,t)} {\partial x}
 + \frac{1}{n(x,t) }
\frac{\partial} {\partial x} j_q (x,t)=-\frac{T_b}{\tau}  \left[\frac{2\epsilon_{int}(x,t)}{\theta(x)}-1 \right],
\ee
which corresponds to Eq. (34) of Chapter II in~\cite{DEGM}, in the case of a system with a single chemical component.
The last term, of course, is not present in~\cite{DEGM} because heat-bath thermostats are not considered there.

Using the continuity equation to eliminate the compression work in the equation for the internal energy we derive
\bea
&&
n(x,t) \left\{ \frac{\partial \epsilon_{int}(x,t)} {\partial t}+ u(x,t) \frac{\partial} {\partial x} \epsilon_{int}(x,t)  -\frac{\pi(x,t)}{n^2(x,t)} \left[\frac{\partial n(x,t)} {\partial t}+ u(x,t) \frac{\partial} {\partial x} n(x,t) \right] \right\}
\nonumber\\
&& =-
\frac{\partial} {\partial x} j_q (x,t)-\frac{T_b}{\tau}  n(x,t) \left[  \frac{2\epsilon_{int}(x,t)}{\theta(x)} -1 \right]
\label{internal}
\eea

In analogy with equilibrium thermodynamics, we can identify the quantity $$\frac{[\ep_{int}(x,t)+ \pi(x,t)/n(x,t)]}{\theta(x)} = s_{h}(x,t)$$ as a good candidate
for the hydrodynamic entropy density and rewrite the last equation as an equation for the entropy.
The first law of thermodynamics $ T d s_{h}= d \ep_{int}+\pi(x,t) d(\frac{1}{n}) $ becomes the local relation
\bea
n(x,t)  \Bigl( \frac{\partial s_{h}(x,t)} {\partial t}+ u(x,t) \frac{\partial} {\partial x} s_{h}(x,t)   \Bigr)
=-
\frac{1}{\theta(x) }   \frac{\partial} {\partial x} j_q (x,t)- \frac{1}{\tau}  \frac{T_b}{\theta(x) }   n(x,t)[ \frac{ 2\epsilon_{int}(x,t)}{\theta(x) }  -1]
\label{entrop}
\eea
which can be also rewritten as
\bea
&&
n(x,t)  \Bigl( \frac{\partial s_{h}(x,t)} {\partial t}+ u(x,t) \frac{\partial} {\partial x} s_{h}(x,t)   \Bigr)
\nonumber\\
&& =- \frac{\partial} {\partial x} \frac{j_q(x,t)}{\theta(x)}
-\frac{1}{\theta^2(x,t) } j_q(x,t)  \frac{\partial} {\partial x} \theta(x)- \frac{1}{\tau}  \frac{T_b}{\theta(x) }   n(x,t)[ \frac{ 2\epsilon_{int}(x,t)}{\theta(x) }  -1 ] .
\label{entropyequation}
\eea
In  eq.\eqref{entropyequation} we identify the {\em internal} entropy flux 
$
j_{s}(x,t) \equiv  \frac{j_q(x,t)}{\theta(x)},
$
which is the difference between the total entropy flux and the convective entropy flux \cite{DEGM} ,
and define the  local entropy production  of the system as:
\be
\sigma_s(x,t)=-\frac{1}{\theta^2(x,t) } j_q(x,t)  \frac{\partial } {\partial x} \theta(x) =   j_q(x,t)  \frac{\partial } {\partial x}\frac{1}{ \theta(x)} .
\label{entroproduction}
\ee

A closed expression for the heat flux  $j_q(x,t)=n(x,t)\langle (v
-u(x,t))^3\rangle/2$ requires the knowledge of the third velocity moment   or alternatively one must use
a phenomenological closure relating heat flux to the temperature gradient, e.g.  $j_q(x,t) =-\kappa(x,t) \nabla \theta(x)$ with some positive thermal conductivity $\kappa$ proportional to the density $n(x,t)$.
In such a case, the local entropy production, Eq. \eqref{entroproduction}, is positive.  Such a phenomenological assumption contrasts with the standard definition
 $j_q(x,t) =-\kappa \nabla T(x,t)$, however  in the limit of small $\tau$ the temperature $T(x,t)$ can be approximated by 
  $\theta(x)$ so that 
  $$\sigma_s(x,t)     =\frac{ \kappa(x,t) }{\theta^2(x,t) }  \Bigl(\frac{\partial } {\partial x} \theta(x)\Bigr )^2 \geq 0 .
  $$
Finally, the local entropy flux towards the surrounding medium (i.e. towards the active bath) reads:
\be
\sigma_m(x,t)= - \frac{1}{\tau}  \frac{T_b}{\theta^2(x) }   n(x,t)
\Bigl[  2\epsilon_{int}(x,t)  - \theta(x) \Bigr].
\ee 
By rewriting it as 
\be
\sigma_m(x,t)= - \frac{1}{\tau}  \frac{T_b}{ \theta^2(x) }   n(x,t)
\Bigl[ T(x,t)  - \theta(x) \Bigr],
\ee 
one can see that $\sigma_m(x,t)$ corresponds (using Eq~\eqref{tildeq}) to  the density of medium entropy production $\dot{\tilde{q}}(x,t)/\theta(x)$ featuring in  the integrand in the r.h.s. of Eq.~\eqref{sm}.


 In conclusion, we can write
 \bea
 \frac{\partial } {\partial t} n(x,t) s_{h}(x,t) +  \frac{\partial} {\partial x} (n(x,t) u(x,t) s_{h}(x,t))    + \frac{\partial} {\partial x}\frac{ j_q(x,t)}{\theta(x)}=\sigma_s(x,t)+\sigma_m(x,t),
\label{entropyequation2}
\eea
where the second and the third term on the l.h.s. together represents the entropy current density, whereas in the r.h.s.
we have the total entropy production density   as the sum  of the   entropy production of the system
and of the medium.  Let us remark that formula \eqref{entropyequation2} is in agreement with standard treatments
of non equilibrium thermodynamics \cite{kreuzer1981nonequilibrium}.
 
\subsubsection*{ UCNA: an approximate treatment of the Fokker-Planck-Kramers equation}
\label{UCNA}
In the present section, we show that it is straightforward to derive the static UCNA equation  in a non perturbative fashion from the hydrodynamic
equations \eqref{brinkman0d}- \eqref{brinkman2d} and a series of approximations. First of all we assume that the kinetic temperature
of the active particles is equal to the temperature $\theta$ of the non uniform heat bath: so that
in equation \eqref{brinkman2d} we put 
\be
T(x,t)=\theta(x)=\frac{T_b}{\Gamma(x)}.
\label{teqtheta}
\ee
 Next, we
 we assume that the l.h.s. of eq. \eqref{brinkman1d}  representing  the hydrodynamic derivative of the average velocity
vanishes, so that:
\begin{equation}
u(x,t) \left( \frac{\partial n(x,t) } {\partial t}+  \frac{\partial} {\partial x} [n (x,t) u(x,t) ]\right) + n(x,t) \left(\frac{\partial u(x,t) } {\partial t} + u(x,t) \frac{\partial} {\partial x} u(x,t)  \right)=0 
\end{equation}
The first parenthesis vanishes by the conservation of the particle number, the second parenthesis is zero when the volume element does not
accelerate, which means that there is a dynamical equilibrium between frictional forces and external forces.
Thus from the momentum equation \eqref{brinkman1d} 
we have the balance condition:
\begin{equation}
n(x,t) u(x,t)= -\frac{\tau}{\Gamma(x)}\Bigl[  \frac{\partial \pi(x,t)} {\partial x} 
-\frac{f(x)}{ \tau \gamma}n(x, t) \Bigr]  =  -\frac{1}{\Gamma(x)}\Bigl[ D \frac{\partial } {\partial x} (\frac{n(x,t)} {\Gamma(x)})
-\frac{f(x)}{ \gamma}n(x, t) \Bigr] 
\end{equation}
where in the last equality we have used the following relation between pressure, density and $\Gamma(x)$:  $\pi(x,t)=D n(x,t)/(\tau \Gamma(x))$.
Notice that we have used the definition of pressure (given after \eqref{q4q4}) :
$$
\pi(x,t)=n(x,t) T(x,t)=n(x,t) \theta(x)=T_b \frac{ n(x,t)}{\Gamma(x)}
$$
and the equality between $\theta$ and $T$.
Such a pressure coincides with the definition of pressure in the case of the static  UCNA \cite{marconi2016pressure}.
Finally, using the continuity equation we eliminate the hydrodynamic velocity and obtain the UCNA equation:
\be
\frac{\partial n (x, t)}{\partial  t}=\frac{\partial} {\partial x} \frac{1}{\Gamma(x)}\Bigl[ D \frac{\partial } {\partial x} (\frac{n(x,t)} {\Gamma(x)})
-\frac{f(x)}{ \gamma}n(x, t) \Bigr] 
\label{ucnarefer}
\ee
which is a modified diffusion equation.
Alternatively,as shown in the appendix one can derive systematically by a multiple-time scale method an equation for the evolution of $n(x,t)$ which is equivalent to  \eqref{ucnarefer} to first  order in $\tau$.

When the momentum current vanishes we have from \eqref{ucnarefer} the following condition:
$$f(x)=D\gamma \frac{\partial }{\partial x}\left(\frac{n(x,t) }{\Gamma(x)} \right)
$$
which is the hydrostatic equation discussed in previous papers by some of us  \cite{marconi2015towards}
showing that the dynamical and the static definitions of active pressure coincides  .
According  to \eqref{entropyequation}, the UCNA,  since $j_q(x,t)=0$ and $T(x,t)=\theta(x)$, corresponds to
\be
 \frac{\partial } {\partial t}[ n(x,t)  s_{h}(x,t)] +\frac{\partial} {\partial x}   [ n(x,t) u(x,t)  s_{h}(x,t)]  =0,
\ee
i.e. it coincides with vanishing entropy production, as previously discussed.

Finally, let us remark that  due to the equality \eqref{teqtheta} the heat flux
$\dot{\tilde q}(x,t)$ defined by  \eqref{tildeq} vanishes everywhere in the UCNA, and not only its integral.
 As underlined by Cates and Nardini \footnote{M.E. Cates, C. Nardini, Colored noise models of active
particles, http://www.condmatjournalclub.org/2690}  the UCNA method  maps the GCN  non-equilibrium description into an equilibrium one
and rules out macroscopic steady-state fluxes.  Hence,  $\sigma_m (x,t)$ vanishes within the UCNA, but 
we do not expect this vanishing to occur in the  GCN case. In order to explicitly  observe a negative value of  the entropy production   it is necessary to consider an approximation
 more refined than the UCNA, a task which will be carried out in the next section by using a perturbative method.

\section*{Results and discussion}

\subsubsection*{A systematic perturbative solution of the Fokker-Planck equation for the GCN}
\label{multscale}


We have seen that
the UCNA equation \eqref{ucnarefer},
 derived  from the exact evolution equation  \eqref{kunob}  for the phase-space distribution function, $P(X,V,\bar t)$
  by eliminating the velocity in favour of the configurational degrees of freedom, satisfies the detailed balance condition. 
   In the appendix, we show how
to construct   a systematic expansion of the Fokker-Planck equation  in the small parameter  $\sqrt \tau\propto 1/\zeta$,  and obtain a time-dependent equation for  the reduced  spatial distribution function and its corrections about the solution with $\tau=0$.
Such an expansion when truncated at the second order in the perturbative parameter $1/\zeta^2$ leads to the same evolution equation for the reduced spatial density  as the one  introduced by Fox \cite{fox1986functional} :
\be
\frac{\partial}{\partial  t} n^{Fox} (x, t)  =\frac{\partial} {\partial x} \Bigl[ \frac{\partial} {\partial x}\Bigl( D^{Fox}(x) n^{Fox}(x,t) \Bigr) - \frac{f(x)}
{ \gamma}n^{Fox}(x,t)  \Bigr] 
\label{foxequation}
\ee
with $D^{Fox}(x) =\frac{D}{\Gamma(x)} $. One can observe that it has the same time independent (and zero flux) solution as the UCNA\eqref{ucnarefer}.
Both the Fox and the UCNA description represent a contracted description 
in terms of a spatial distribution with respect to the phase-space distribution of the  GCN,
which is fully described by  Kramers' equation.  The Fox approximation emerges in the limit of small $\tau$,
or small P\'eclet number, when for small $\tau$ the velocity distribution thermalises rapidly and  reaches 
a gaussian shape. The method that we shall discuss below gives quantitative support to this intuitive idea.
If $\tau$ (or P\'eclet) is small the particles lose memory of their initial
velocities after a time span which is of the order 
of the time constant $\tau$ so that the velocity distribution 
soon becomes stationary. While  the derivation of the UCNA equation  can be done 
simply eliminating the time derivative of the  highest order
 in the stochastic differential
equation and then writing the associated equation for the distribution function of positions only,
such a procedure does not tell us anything about the deviations of the velocity distribution
from a local Maxwellian. In order to study the corrections,
 it is convenient to consider a general method of kinetic theory which allows extracting  a reduced description  from a finer one. An instance of such a method is the so-called Hilbert-Chapman-Enskog approach which allows  deriving,
starting from the  transport equation for the phase space distribution, the  Navier-Stokes equations
under the form of  a series expansion with the Knudsen number as the perturbation
parameter ( and to successive orders  in Knudsen number the Burnett and super-Burnett equations)\cite{chapman1970mathematical}.
In the overdamped case, the application of the Hilbert approach is even simpler because there is only one conserved, slow mode, namely the diffusive density mode.  
The remaining momentum and energy variables are slaved to the density,
so that one can reduce the Kramers equation  to a Smoluchowski-like equation involving only the
density \cite{marconi2010dynamic}. The reduction is achieved by the multiple-time scale method as illustrated in the appendix where we report the
necessary details  of the calculation. In this section, we use some results concerning the 
steady state phase-space distribution function.  

\subsubsection*{ Mean square velocity and entropy production beyond the order $1/\zeta^2$}
In order to observe violations of the detailed balance condition we shall consider the fourth order in the $1/\zeta$ expansion
of the phase space distribution function.
To this purpose, one quantity of capital interest to compute the heat-flux and the entropy production is the mean square
velocity.  Using the expansion   \eqref{hermiteexpansion2} of the steady state $P(X,V)$ given in the appendix  up to fourth order in $\zeta^{-1}$ we have the formula for $\langle V^2 \rangle_X$, the dimensionless mean squared velocity at given position, which is also the local kinetic temperature of the particle:
\be
\langle V^2 \rangle_X\equiv 
\frac{\int dV V^2 P(X,V)}{ \int dV P(X,V)} \approx 1+\frac{2}{\zeta^2}\frac{\psi_{22}(X)} {\psi_{00}(X)}   +\frac{2}{\zeta^4}\frac{\psi_{42}(X)} {\psi_{00}(X)}   ,
\label{v2formula}
\ee
where $\psi_{00}(X)= \int dV P(X,V)$ is the marginalized distribution function and   $ \int dV V^2 P(X,V)=\psi_{00}(X)+2\psi_{22}(X)/\zeta^2+2\psi_{42}(X)/\zeta^4 + 
O(\zeta^{-6})$ according to \eqref{hermiteexpansion2}.
Substituting eq. \eqref{psi22} and \eqref{psiquarto}
we obtain the following approximation:
\be
\langle V^2 \rangle_X    = 1-\frac{1}{\zeta^2} \frac{d^2 U}{d X^2}+\frac{1}{\zeta^4} (\frac{d^2 U}{d X^2})^2 +\frac{1}{\zeta^4} R(X) ,
\label{v2series}
\ee
where the remainder, $R(X)$ is found using the result   \eqref{psiquarto}:
\be
R(X)= - \frac{3 }{2} [\frac{d^4 U}{d X^4}   - \frac{d U}{d X} \frac{d^3 U}{d X^3} ]   .
\label{remainder}
\ee
We compare now the expression \eqref{v2series} with the approximate UCNA prediction:
\be
\langle V^2 \rangle_X^{ucna}   =\frac{1}{g(X)} \approx 1-\frac{1}{\zeta^2} \frac{d^2 U}{d X^2} + \frac{1}{\zeta^4} (\frac{d^2 U}{d X^2})^2 
\label{tucna}
\ee
and conclude that
 up to order $\zeta^{-2}$ the result \eqref{v2series} of the multiple-time scale method and of the UCNA agree and give the same  value the kinetic temperature $<V^2>_X$. On the other hand, at order  $\zeta^{-4}$
 the  two formulae are identical  only when $R(X)=0$, a situation occurring both in the case of a particle confined to an harmonic potential well or in a constant force field, where the entropy production vanishes, in agreement with the fact that the detailed balance holds.

 Let us consider the difference, $\Delta(X)$,  between the local temperature of the bath $1/g(X)$ and the kinetic temperature  $\langle V^2(X)\rangle_X$ whose sign controls the direction of the local heat exchange with the bath (positive values correspond to an heat flux towards the particle, whereas for negative values the particle transfers heat to the bath): 
 \be
 \Delta(X)=\frac{1}{g(X)}-\langle V^2(X)\rangle_X=  \langle V^2(X)\rangle_X^{ucna}-  \langle V^2(X)\rangle_X \approx - \frac{R(X)}{\zeta^4}
 \ee
 where the last approximate equality follows from \eqref{v2series}.
  Thus inserting this result in eq. \eqref{essem} we find:
 \be
 \langle \dot S\rangle_m= \zeta\int dX g^2(X) \Delta(X) \psi_{00}(X)=-\frac{1}{\zeta^3}\int dX g^2(X) R(X)
\psi_{00}(X) .
\label{smdelta}
 \ee
Using the scaling $\bar t= \frac{t}{\sqrt{\tau}} \sqrt{D}/l$
we can see that the lowest order estimate of the dimensional entropy production rate $\dot s_m(t) \sim \tau^2$ 
in agreement with Fodor et al. \cite{fodor2016far} .

\subsubsection*{A numerical example }
\label{example}

In order to illustrate the  above theoretical  results, we have performed some numerical simulations
and integrated the stochastic eqs.~\eqref{sistema2} by an Euler algorithm optimized for GPU execution~\cite{maggi2015multidimensional}.
We focus on the archetypical case of the double well potential: $w(x)=x^4/4-x^2/2$. This potential has been used as the simplest model for testing the colored noise approximation schemes~\cite{hanggi1995colored} and it has been recently considered as a Ginzburg-Landau potential for the phase separation of attractive active particles~\cite{paoluzzi2016critical}.  
For this example we set $\gamma=1$, $D=1$ and $\tau=0.7$, we use a time-step $dt=10^{-3}\tau$ and we average 1024 independent trajectories staring in $x=0$ for more than $10^7$ steps. In Fig.~1(a) we report the full phase-space distribution  $p(x,v)$, the corresponding $p(x)$ is shown in Fig.~1(b). 
In Fig.~1(c) we show the ``temperature" profiles $\langle v^2 \rangle_x$ and $\theta(x)$ determining the local heat flow $\dot{\tilde{q}}(x)$ (eq.~(\ref{tildeq})) and the local entropy production $\dot{\tilde{q}}(x)/\theta(x)$ (see (\ref{sm})) shown in Fig.~1(d). We find in agreement with the theory of section Model and Methods  that in the stationary state $\int dx \dot{\tilde{q}}(x) \approx 0$ as shown in Fig.~1(d) the positive lobe of $\dot{\tilde{q}}(x)$ is compensated by two smaller negative lobes. Differently for the entropy we have $\int dx \dot{\tilde{q}}(x)/\theta(x) \approx -0.34<0$ in agreement with inequality (\ref{gencla}) as also shown 
in Fig.~1(d) where $\dot{\tilde{q}}(x)/\theta(x)$ has much more pronounced negative lobes (dashed line, gray area).
 Let us remark that these numerical findings are in agreement with our theoretical formula \eqref{deltax}
 for the velocity difference.
 
Finally, we can compare these results with the
prediction of the theory of the previous subsection, use the quartic potential $X^4/4-X^2/2$ and employ \eqref{remainder} to evaluate the mean square velocity
\be
\langle V^2 \rangle_X    = (1+\frac{1}{\zeta^2}- \frac{9}{\zeta^4})- 3 X^2 (\frac{1}{\zeta^2}+\frac{5}{\zeta^4})
+ X^4  \frac{18}{\zeta^4}
\label{v2seriesb}
\ee
 and observe that  in this case $\langle V^2 \rangle_X$ differs from $1/g(X)$ (the non dimensional equivalent of $\theta(x)$) by a quantity $-\Delta(X)$, with  :
\be
\Delta(X)=\frac{9}{\zeta^4} (1+X^2-X^4) \,
\label{deltax}
\ee
which  assumes positive values near the origin, but is negative for large values of $X$ as also shown by the numerical solution
Fig.~1(c) . The negativity of $\Delta(X)$ for large $X$  can  explain  why the entropy production predicted by eq. \eqref{smdelta}
 is negative in agreement with the numerical result of Fig.~1(d). 
 
\begin{figure}[h]
\includegraphics[width=.85\textwidth] {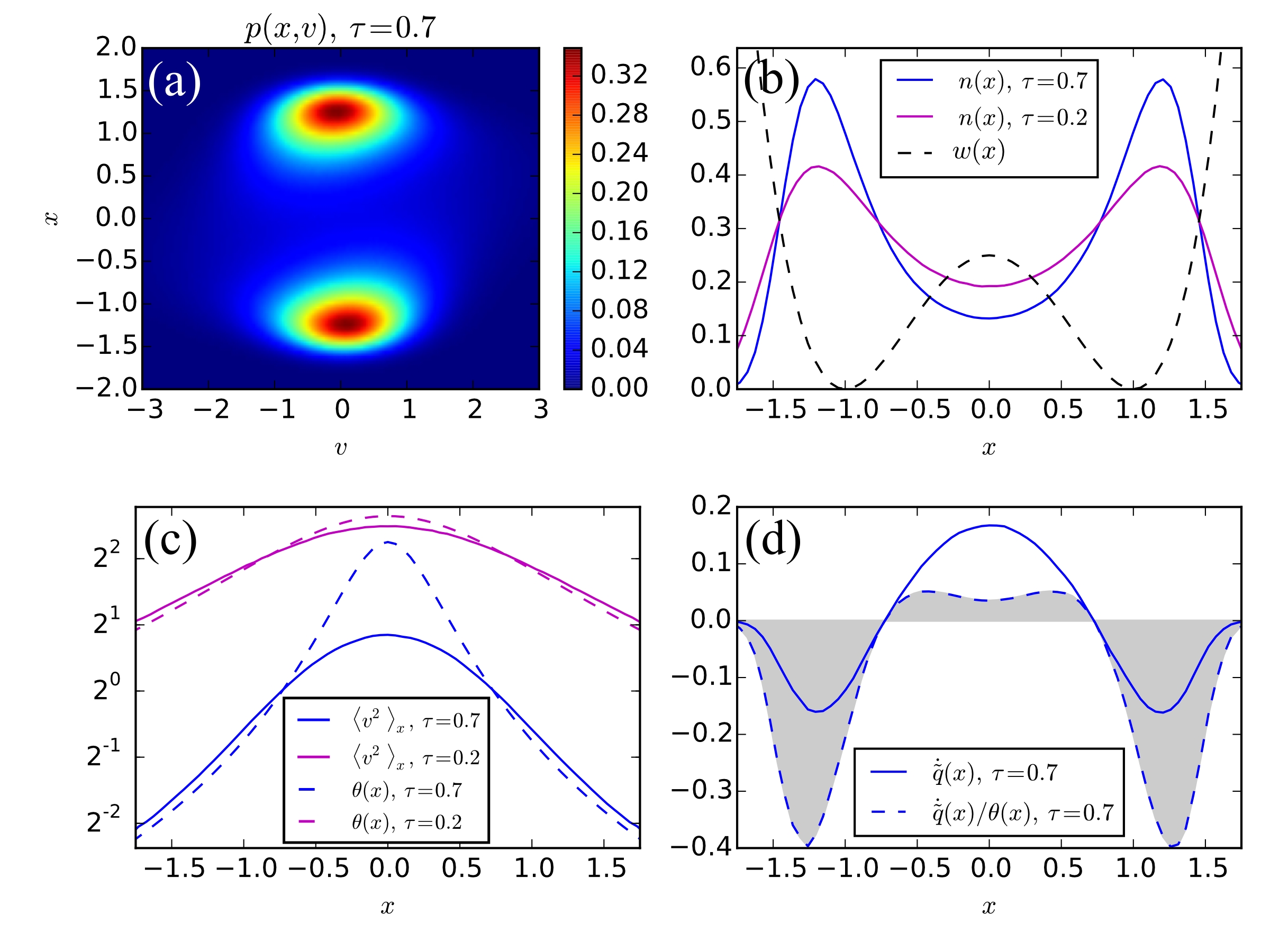}\label{fig:f1} 
\caption{(\textbf{a}) Stationary probability distribution $p(x,v)$, obtained numerically in the case of a double-well potential
and  persistence time $\tau=0.7$.
(\textbf{b}) Position probability distribution $n(x)$ (full line) obtained numerically for the potential $w(x)$ (dashed line)
for two different values of the persistence time (blue $\tau=0.7$ and magenta $\tau=0.2$).
The potential is shifted upwards by an inessential constant $1/4$ for reasons of presentation.  
(\textbf{c}) ``Temperature" profiles $\langle v^2  \rangle_x$ and $\theta(x)$ (full and dashed lines respectively and blue $\tau=0.7$ and magenta $\tau=0.2$).
In the case $\tau=0.7$,
notice the crossover of the difference  $\theta(x)-\langle v^2  \rangle_x$ from positive values at small values of $x$ to negative values
at larger values of the coordinate. When $\tau=0.2$, the difference is very small.
(\textbf{d})
Local heat flow and local entropy production 
(full and dashed lines respectively) for persistence time $\tau=0.7$.
Both quantities are negative in the potential wells
since there the particle transfers heat to the bath, whereas the opposite occurs in the peak region.}
Note that the integral of the entropy is negative as evidenced by the grey area, in agreement with (\ref{gencla}).
\end{figure}

\section*{Summary and Conclusion}

We have analysed the energetics and thermodynamics of a stochastic
model for active particles, in the non-interacting case and restricted
to a single dimension. Our results show that the active bath (the
fluctuating active force $a(t)$) can be interpreted as at local
equilibrium at temperature $\theta(x)=T_b/\Gamma(x)=D/[\tau
  \Gamma(x)]$, with $D$ the active diffusivity, $\tau$ the active
persistence and $\Gamma(x)$ a renormalisation factor which depends
on the external force field. As a matter of fact, the particle is at
equilibrium with the active bath (yielding a zero entropy production)
only when the force field is flat, i.e. $\Gamma(x)=1$. Otherwise,
there is an active entropy production which - in the stationary state
- is eliminated as entropy flux to the medium, taking the simple
expression Eq.~\eqref{sm} and obeying a generalised Clausius relation,
Eq.~\eqref{gencla}. By observing the system on a longer timescale
(UCNA approximation), the discrepancy between local temperature and
$\theta(x)$ becomes negligible and the particle appears as at
equilibrium with the active bath. A hydrodynamic approach which
describes the evolution of density, velocity and temperature fields,
allows one to define local internal energy, local entropy and their
local ``thermodynamic'' relations. The result is consistent with the
global picture obtained from the analysis of the Kramers equation.  In
the hydrodynamic description, the UCNA approximation is interpreted as
the analog of the Euler hydrodynamic solution of the Boltzmann
equation. There one employs a local Maxwell-Boltzmann distribution and
obtains an approximate solution of the transport equation and a set of
conditions which determine the local values of the density, fluid
velocity and temperature, the hydrodynamic fields. In the present case,
the solution of the Kramers equation is a local Maxwellian at
temperature $1/g(X)$ and density $\pi(X)$ given by the UCNA.
Finally, we have discussed how to  improve the theory and obtain 
a non vanishing entropy production by deriving a systematic expansion of the 
phase-space distribution function in powers of $\sqrt \tau$ without invoking the detailed balance condition.
When this condition is violated we observe numerically and theoretically a negative entropy production
 and a dependence of the local mean square velocity
 on position.
 
 As far as approximations, such as the UCNA, involving the detailed balance condition, are concerned
 their predictions  about the steady state structure of many-particle systems should be valid up to order 
 $1/\zeta^2$, i.e. order $\tau$.  In this case, the application of the concept of effective potential can lead to simple treatments
 of non uniform systems but are not reliable when the activity becomes large and higher order terms in the perturbative expansion are important. 

A generalisation of the above discussion to many dimensions and to
interacting particles is expected to give similar results, but
certainly, deserves further investigation. A promising line of research
is an experimental verification of the generalised Clausius
inequality, Eq.~\eqref{gencla}.


\section*{Appendix A}
\label{appb}

 \subsubsection*{Energetics of the hydrodynamic equations}

A well-known procedure to derive
from the transport equation the coupled evolution equations for the
density, the momentum density, and the average kinetic energy is to
project the the transport equation over the subspace $(1,V, V^2)$, that is multiply
eq.\eqref{kramers0b} by these functions and integrate over $V$.  This
method leads to the first three members of an infinite hierarchy of
equations for the velocity moments.  Such equations are nothing else
as the balance equations for number density, momentum density, and
kinetic energy density.

 \bea && \frac{\partial \rho (X,\bar
  t)}{\partial \bar t}+ \frac{\partial} {\partial X} J_v (X,\bar t)=0
\label{brinkman0}
\\
&&
\frac{\partial J_v(X,\bar t)} {\partial \bar t}  + 2\frac{\partial K(X,t)} {\partial X} 
-F(X)\rho(X,\bar t)=-\zeta g(X) J_v(X,\bar t)
\label{brinkman1}
\\
&&
\frac{\partial K(X,\bar t)} {\partial \bar t}  + \frac{\partial J_k(X,t)} {\partial X} 
-F(X) J_v (X,\bar t)=-2\zeta g(X) K(X,\bar t)+\zeta  \rho(X,\bar t) \, ,
\label{brinkman2}
\eea
where we have introduced the number density $\rho(X,\bar t)=\int dV P(X,V,\bar t)$, the
momentum density 
$J_v(X,\bar t)=\int dV V P(X,V,\bar t)$, the
kinetic energy density
$K(X,\bar t) = 1/2\int dV V^2 P(X,V,\tau)$  
and kinetic energy flux  $J_k(X,\bar t)=1/2\int dV V^3 P(X,V,\bar t)$.
 Notice that the presence of the last term  in eq. \eqref{brinkman2}
tends to maintain the fluid at local temperature $1/g(X)$ (which in dimensional form corresponds to the temperature $\theta(x)$.

If the hierarchy of moment equations is truncated,
by supplementing the constitutive equations, 
one recovers the analog of hydrodynamic equations with
dissipation. 

Let us integrate  Eqs.~\eqref{brinkman1} and ~\eqref{brinkman2} over whole phase space (so that the integral of spatial  gradients vanish). We get
 \be
\frac{d}{d\bar t}  \int dX   J_v(X,\bar t)
=  \int dX  \Bigl [ F(X)\rho(X,\bar t) -\zeta g(X) J_v(X,\bar t) \Bigr]
 \label{firsteq}
 \ee
 
 \be
   \frac{d}{d\bar t}  \int dX  K(X,\bar t) = 
 \int dX   \Bigl [F(X) J_v (X,\bar t)-2\zeta g(X) K(X,\bar t)+\zeta  \rho(X,\bar t) \Bigr]
 \label{secondeq} 
 \ee
 If the particle (described by the current $J_v$)  does not accelerate the l.h.s. of eq. \eqref{firsteq} vanishes:
 \be
 \int dX   F(X)\rho(X,\bar t) =  \zeta \int dX g(X) J_v(X,\bar t),
 \ee
 and  if the space averaged  kinetic energy, represented by the l.h.s.of \eqref{secondeq} does not change , the work done per unit time by the external force (power) 
 equals the frictional dissipation
 \be
 \int dX  F(X) J_v (X,\bar t)= \zeta\int dX  \Bigl [ 2 g(X) K(X,\bar t)-  \rho(X,\bar t) \Bigr]. 
 \label{q4q4}
 \ee

\section*{Appendix B}
\label{multipletimescale}

\subsubsection*{Multiple time-scale analysis of the Kramers equation}

In this appendix, we derive the Fox equation for the distribution
function \cite{fox1986functional,fodor2016far} by a multiple
time-scale analysis following the same method employed in
ref. \cite{marconi2006nonequilibrium,marini2007}. It allows deriving
in a systematic fashion the configurational Smoluchowski equation from
the Kramers equation via the elimination of the velocity degrees of
freedom~\footnote{ The multiple time scale method presented here is different from the one we recently reported because
to perform the perturbative calculation  we employ an
Hermite uniform basis in this appendix \cite{marconi2016effective}  }.

In order to perform the multiple-time scale    according to  the method of Titulaer \cite{titulaer1978systematic}, we rewrite as eq. \eqref{kramers0b}
using the operator  
$
L_{FP} =\frac{\partial}{\partial V}\Bigl[
\frac{\partial }{\partial V }+ V\Bigl]  
\label{fokkerpb}
$:
\begin{equation} 
\frac{\partial  P(X,V,\bar t)}{\partial \bar t} +V \frac{\partial }{\partial X}  P(X,V,\bar t) 
+ F(X,\bar t) \frac{\partial }{\partial V}  P(X,V,\bar t)+\frac{1}{\zeta} \frac{\partial F}{\partial X} \frac{\partial }{\partial V} V  P(X,V,\bar t)
=\zeta L_{FP} P(X,V,\bar t) .
\label{kramers0bb}
\end{equation}
The idea, is to use the non-dimensional parameter $\zeta^{-1}$, which  vanishes when $\tau\to 0$ at fixed $v_T$, to perform a perturbative expansion  by projecting eq. \eqref{kramers0bb} onto the Hermite eigenfunctions       $ H_{\nu}(V)
= (-1)^{\nu}   \frac{1}{\sqrt{2\pi}} 
 \frac{\partial^{\nu}}{\partial V^{\nu}} \exp(-\frac{1}{2}V^2) .
$ of of $L_{FP}$ corresponding to
 the eigenvalues  $\nu=0,-1,-2 , .., -\nu $.
Since the Hermite basis is complete and orthogonal, one may represent the solution $P(X,V,\bar t)$ as a linear superposition
of $H_\nu(V)$ with time and space dependent coefficients $\phi_{\nu}(X,\bar t)$
\begin{equation}
 P(X,V,\bar t) \equiv \sum_{\nu=0}^{\infty}\phi_{\nu}(X,\bar t) H_{\nu}(V).
\label{expansion}
\end{equation}
We shall prove that 
to leading order in $\zeta^{-1}$ the time-dependent coefficients in \eqref{expansion} with $\nu>0$ vanish
as $e^{-\nu \zeta \bar t}$, accounting for the fast relaxation of the momentum, kinetic energy and higher velocity moments of the 
distribution $P$, whereas the mode with $\nu=0$ displays the slowest decay and after a transient
 of the order
$\zeta^{-1}$ will provide the largest contribution to the series  \eqref{expansion} .
The machinery to construct such a solution employs
slow and fast time-scale  variables, $\bar t_0,\bar t_1,\bar t_2,...$  replacing the original  the time variable, $\bar t$. They are
 related to the original variable
by  $\bar t_n=\zeta^{-n}\bar t$ and treated as if they were independent.  
Thus, the physical
time-dependent function, $ P(X,V,\bar t)$,  
is replaced by an auxiliary function, $ P_a(X,V,\bar t_0,\bar t_1,..)$,  
that depends on all  $\bar t_n$.
By using a perturbation theory in powers of $\zeta^{-1}$ illustrated below one constructs order by order the solution, i.e.
determines the coefficients of the series \eqref{expansion} in terms of the coefficients of lower order in $\zeta^{-1}$ and
 once the coefficients  corresponding to the various orders have been 
determined, one returns to the original time variable and to the
original distribution.
One begins by replacing the time derivative
with respect to $\bar t$ by a sum of partial derivatives:
\begin{equation}
\frac{\partial}{\partial \bar t}=\frac{\partial}{\partial \bar t_0}
+\frac{1}{\zeta} \frac{\partial}{\partial \bar t_1}
+\frac{1}{\zeta^2} \frac{\partial}{\partial \bar t_2}+..
\label{mult}
\end{equation}
and expands  the coefficients $\phi_\nu$   in powers of $\zeta^{-1}$ 
\begin{equation} 
 P_a(X,V,\bar t_0,\bar t_1,\bar t_2,..)=
\sum_{s=0}^{\infty} \frac{1}{\zeta^s} 
\sum_{\nu=0}^{\infty}  
\psi_{s \nu}(X,\bar t_0,\bar t_1,\bar t_2,..)  H_{\nu}(V) .
\label{pn}
\end{equation}

One, now, substitutes the time derivative~(\ref{mult})
and expression~\eqref{pn} into eq. 
(\ref{kramers0bb}) 
and equates terms 
of the same order in $\zeta^{-1}$.  This procedure gives a hierarchy of relations
between the amplitudes $\psi_{s \nu}$, allowing to express those of order $s>0$ in terms of $\psi_{00}$.
To order $\zeta^0$ eq. \eqref{kramers0bb}  gives:
\begin{equation}
L_{FP} \Bigr[\sum_\nu \psi_{0\nu}  H_\nu\Bigr]=0
\label{g0}
\end{equation} 
which shows that only the amplitude $\psi_{00}$, associated with the null eigenvalue ($\nu=0$), is non vanishing.
Next, we consider terms of order  $\zeta^{-1}$ and obtain:
\begin{equation}
L_{FP}  \sum_{\nu>0}\Bigr[ \psi_{1\nu} H_\nu \Bigr]=
\frac{\partial \psi_{00}}{\partial \bar t_0}  H_0+
\Bigl(V \frac{\partial }{\partial X} 
+ F \frac{\partial }{\partial V}\Bigl)
 H_0 \psi_{00}\label{g1}
\end{equation} 
After some straightforward calculations and
 equating the coefficients multiplying the same Hermite polynomial
we find:
\begin{equation}
\frac{\partial \psi_{00}}{\partial \bar t_0}=0
\label{psi0ta}
\end{equation} 
$\psi_{1\nu}=0$ when $\nu>1$ and
\begin{equation}
\psi_{11}=-D_X \psi_{00}
\label{psi0t}
\end{equation} 
with $D_X\equiv \Bigl( \frac{\partial }{\partial X }-F \Bigl)$.
According to \eqref{psi0ta}
the amplitude $\psi_{00}$ is constant with respect to  $\bar t_0$
and so is $\psi_{11}$
being a functional of $\psi_{00}$.
The equations of order $\zeta^{-2}$ give 
the   following conditions:
\bea
&&\frac{\partial \psi_{00}}{\partial \bar t_1}=
-  \frac{\partial }{\partial X} \psi_{11} = \frac{\partial }{\partial X} D_X \psi_{00} \\
\label{psi0t1}
&&
\psi_{2\nu}=0 \quad  \nu\neq 2 \\
&&
 \psi_{22}=-\frac{1}{2}[D_X \psi_{11}- \frac{d F}{d X} \psi_{00}]= \frac{1}{2} \frac{d F}{d X} \psi_{00}
 \label{psi22}
\eea
The third order  order, $\zeta^{-3}$, equations give: 
\bea
&&\frac{\partial \psi_{00}}{\partial \bar t_2}=0 \\
\label{psi0t1b}
&&
 \psi_{31}= -   \frac{\partial }{\partial X} [  \frac{d F}{d X}  \psi_{00} ]   =- [F\frac{d F}{d X}   + \frac{d^2 F}{d X^2}] \psi_{00}
 \label{psi31}\\
 &&
 \psi_{32}=0\\
&&
\psi_{33}=-\frac{1}{3}[D_X \psi_{22}  -\frac{d F}{d X} \psi_{11} ] =-\frac{1}{6}  \frac{d^2 F}{d X^2}  \psi_{00} \label{psi33} \\
\eea
Finally the order $\zeta^{-4}$ of the expansion yields 
\be
\frac{\partial \psi_{00}}{\partial \bar t_3}=-  \frac{\partial }{\partial X} \psi_{31}= 
  \frac{\partial^2 }{\partial X^2} [  \frac{d F}{d X}  \psi_{00}   ] 
\ee
and
\be
 \psi_{42}=-\frac{1}{2} D_X \psi_{31}-  \frac{3}{2}    \frac{\partial }{\partial X}   \psi_{33} =\frac{1}{4}\psi_{00}
 \Bigl[ 2(\frac{d F}{d X})^2+ 3\frac{d^3 F}{d X^3}   +3 F  \frac{d^2 F}{d X^2} 
 \Bigr]
 \label{psiquarto}
\ee
Putting together results \eqref{psi0t} and \eqref{psi31} and restoring the original time variable $\bar t$ we obtain at the following closed equation for $\psi_{00}$:
\be
\frac{\partial \psi_{00}(X,\bar t)}{\partial \bar t}=\frac{1}{\zeta}  \frac{\partial }{\partial X} \Bigl[  \Bigl( \frac{\partial }{\partial X }-F \Bigl) \psi_{00}+\frac{1}{\zeta^2}  \frac{\partial }{\partial X}  \Bigl( \frac{d F}{ d X} \psi_{00} \Bigr) \Bigr]
\ee

We, now, compare such a result with Fox's result at the same order in the perturbation parameter:
let us write Fox equation \eqref{foxequation} in non dimensional form as
\be
\frac{\partial \psi_{00}^{Fox}(X,\bar t)}{\partial \bar t}=\frac{1}{\zeta}  \frac{\partial }{\partial X} \Bigl[  \frac{\partial }{\partial X}\frac{  \psi_{00}^{Fox}}{ 1-\frac{1}{\zeta^2}\frac{d F}{d X} }-F  \psi_{00}^{Fox}   \Bigr]\approx \frac{1}{\zeta}  \frac{\partial }{\partial X} \Bigl[  \Bigl( \frac{\partial }{\partial X }-F \Bigl) \psi_{00}^{Fox} +
\frac{1}{\zeta^2}   \frac{\partial }{\partial X}   \Bigl (\frac{d F}{d X}   \psi_{00}^{Fox} \Bigr) \Bigr]
\label{psi00}
\ee
where the last approximate equality follows from taking the large $\zeta$ limit  (i.e. small $\tau$)  and reproduces the result of the  multiple time scale expansion.
Solving eq. \eqref{psi00} in the steady state  we obtain:
$$
\psi_{00}(X) \approx A \exp \Bigl(-U(X)- \frac{1}{2\zeta^2}  (\frac{d U}{d X})^2\Bigr)  \,(1+\frac{1}{\zeta^2} \frac{d^2 U}{d X^2}) .
$$
Moreover, we have found that the phase space distribution, even when the momentum current vanishes,
contains terms proportional to $\bar H_2(V), \bar H_3(V)...$ thus showing the non equilibrium nature of the 
state. To see that, let us write
\bea
&&
 P(X,V,\bar t) \equiv \sum_{\nu=0}^{\infty}\phi_{\nu}(X,\bar t) H_{\nu}(V)
\nonumber\\
&=&\psi_{00}(X,\bar t)  H_0(V)+
[\frac{1}{\zeta }\psi_{11}(X,\bar t)  +\frac{1}{\zeta^3 }\psi_{31}(X,\bar t)  ]H_1(V)+
\frac{1}{\zeta^2 }\psi_{22}(X,\bar t)  H_2(V)+ \frac{1}{\zeta^3 }\psi_{33}(X,\bar t)  H_3(V)+\dots
\label{hermiteexpansion2}
\eea

Thus, the multiple-time scale method gives information about the moments with $\nu>0$ which are 
not considered in UCNA and in Fox's theories  since these only deal with the configurational part of the distribution function.
The coefficients $\psi_{11}, \psi_{31}, \psi_{22},\psi_{33}$ can be computed from  eqs.  \eqref{psi0t}, \eqref{psi31},\eqref{psi33}.
Notice that even in the steady state, the velocity  cumulants of order $\nu\geq2$ do not vanish, showing that the solution does not converge to a Maxwell-Boltzmann distribution corresponding to $\psi_{00}\neq 0$ and to the vanishing of all remaining coefficients $\psi_{\nu k}$.

\section**{Acknowledgments}
U.M.B.M. thanks Luca Cerino for illuminating discussions.
C. Maggi acknowledges support from the European Research Council under the European Union's Seventh Framework programme
(FP7/2007-2013)/ERC Grant agreement  No. 307940.



\begin{thebibliography}{10}
\expandafter\ifx\csname url\endcsname\relax
  \def\url#1{\texttt{#1}}\fi
\expandafter\ifx\csname urlprefix\endcsname\relax\def\urlprefix{URL }\fi
\providecommand{\bibinfo}[2]{#2}
\providecommand{\eprint}[2][]{\url{#2}}

\bibitem{ramaswamy2010mechanics}
\bibinfo{author}{Ramaswamy, S.}
\newblock \bibinfo{title}{The mechanics and statistics of active matter}.
\newblock \emph{\bibinfo{journal}{Annu. Rev. Condens. Matter Phys.}}
  \textbf{\bibinfo{volume}{1}}, \bibinfo{pages}{323} (\bibinfo{year}{2010}).

\bibitem{marchetti2013hydrodynamics}
\bibinfo{author}{Marchetti, M.} \emph{et~al.}
\newblock \bibinfo{title}{Hydrodynamics of soft active matter}.
\newblock \emph{\bibinfo{journal}{Reviews of Modern Physics}}
  \textbf{\bibinfo{volume}{85}}, \bibinfo{pages}{1143} (\bibinfo{year}{2013}).

\bibitem{bechinger2016active}
\bibinfo{author}{Bechinger, C.}, \bibinfo{author}{Di~Leonardo, R.},
  \bibinfo{author}{Lowen, H.}, \bibinfo{author}{Reichhardt, C.} \&
  \bibinfo{author}{Volpe, G.}
\newblock \bibinfo{title}{Active particles in complex and crowded
  environments}.
\newblock \emph{\bibinfo{journal}{Rev. Mod. Phys.}}  (\bibinfo{year}{2016}).

\bibitem{S16}
\bibinfo{author}{Speck, T.}
\newblock \bibinfo{title}{Stochastic thermodynamics for active matter}.
\newblock \emph{\bibinfo{journal}{Europhys. Lett.}}
  \textbf{\bibinfo{volume}{114}}, \bibinfo{pages}{30006}
  (\bibinfo{year}{2016}).

\bibitem{berg2004coli}
\bibinfo{author}{Berg, H.~C.}
\newblock \emph{\bibinfo{title}{E. coli in Motion}}
  (\bibinfo{publisher}{Springer Science \& Business Media},
  \bibinfo{year}{2004}).

\bibitem{tailleur2008statistical}
\bibinfo{author}{Tailleur, J.} \& \bibinfo{author}{Cates, M.}
\newblock \bibinfo{title}{Statistical mechanics of interacting run-and-tumble
  bacteria}.
\newblock \emph{\bibinfo{journal}{Physical Review Letters}}
  \textbf{\bibinfo{volume}{100}}, \bibinfo{pages}{218103}
  (\bibinfo{year}{2008}).

\bibitem{romanczuk2012active}
\bibinfo{author}{Romanczuk, P.}, \bibinfo{author}{B{\"a}r, M.},
  \bibinfo{author}{Ebeling, W.}, \bibinfo{author}{Lindner, B.} \&
  \bibinfo{author}{Schimansky-Geier, L.}
\newblock \bibinfo{title}{Active brownian particles}.
\newblock \emph{\bibinfo{journal}{The European Physical Journal Special
  Topics}} \textbf{\bibinfo{volume}{202}}, \bibinfo{pages}{1--162}
  (\bibinfo{year}{2012}).

\bibitem{stenhammar2014phase}
\bibinfo{author}{Stenhammar, J.}, \bibinfo{author}{Marenduzzo, D.},
  \bibinfo{author}{Allen, R.~J.} \& \bibinfo{author}{Cates, M.~E.}
\newblock \bibinfo{title}{Phase behaviour of active brownian particles: the
  role of dimensionality}.
\newblock \emph{\bibinfo{journal}{Soft Matter}} \textbf{\bibinfo{volume}{10}},
  \bibinfo{pages}{1489--1499} (\bibinfo{year}{2014}).

\bibitem{cates2013active}
\bibinfo{author}{Cates, M.} \& \bibinfo{author}{Tailleur, J.}
\newblock \bibinfo{title}{When are active brownian particles and run-and-tumble
  particles equivalent? consequences for motility-induced phase separation}.
\newblock \emph{\bibinfo{journal}{EPL (Europhysics Letters)}}
  \textbf{\bibinfo{volume}{101}}, \bibinfo{pages}{20010}
  (\bibinfo{year}{2013}).

\bibitem{fily2012athermal}
\bibinfo{author}{Fily, Y.} \& \bibinfo{author}{Marchetti, M.~C.}
\newblock \bibinfo{title}{Athermal phase separation of self-propelled particles
  with no alignment}.
\newblock \emph{\bibinfo{journal}{Physical review letters}}
  \textbf{\bibinfo{volume}{108}}, \bibinfo{pages}{235702}
  (\bibinfo{year}{2012}).

\bibitem{farage2015effective}
\bibinfo{author}{Farage, T.}, \bibinfo{author}{Krinninger, P.} \&
  \bibinfo{author}{Brader, J.}
\newblock \bibinfo{title}{Effective interactions in active brownian
  suspensions}.
\newblock \emph{\bibinfo{journal}{Physical Review E}}
  \textbf{\bibinfo{volume}{91}}, \bibinfo{pages}{042310}
  (\bibinfo{year}{2015}).

\bibitem{peruani2007self}
\bibinfo{author}{Peruani, F.} \& \bibinfo{author}{Morelli, L.~G.}
\newblock \bibinfo{title}{Self-propelled particles with fluctuating speed and
  direction of motion in two dimensions}.
\newblock \emph{\bibinfo{journal}{Physical review letters}}
  \textbf{\bibinfo{volume}{99}}, \bibinfo{pages}{010602}
  (\bibinfo{year}{2007}).

\bibitem{marconi2015towards}
\bibinfo{author}{Marconi, U. M.~B.} \& \bibinfo{author}{Maggi, C.}
\newblock \bibinfo{title}{Towards a statistical mechanical theory of active
  fluids}.
\newblock \emph{\bibinfo{journal}{Soft matter}} \textbf{\bibinfo{volume}{11}},
  \bibinfo{pages}{8768--8781} (\bibinfo{year}{2015}).

\bibitem{marconi2015velocity}
\bibinfo{author}{Marconi, U. M.~B.}, \bibinfo{author}{Gnan, N.},
  \bibinfo{author}{Maggi, M. P.~C.} \& \bibinfo{author}{Di~Leonardo, R.}
\newblock \bibinfo{title}{Velocity distribution in active particles systems}.
\newblock \emph{\bibinfo{journal}{Scientific Reports}}
  \textbf{\bibinfo{volume}{6}}, \bibinfo{pages}{23297 EP}
  (\bibinfo{year}{2016}).

\bibitem{sekimoto2010stochastic}
\bibinfo{author}{Sekimoto, K.}
\newblock \emph{\bibinfo{title}{Stochastic energetics}}, vol.
  \bibinfo{volume}{799} (\bibinfo{publisher}{Springer}, \bibinfo{year}{2010}).

\bibitem{BPRV08}
\bibinfo{author}{Marconi, U. M.~B.}, \bibinfo{author}{Puglisi, A.},
  \bibinfo{author}{Rondoni, L.} \& \bibinfo{author}{Vulpiani, A.}
\newblock \bibinfo{title}{Fluctuation-dissipation: Response theory in
  statistical physics}.
\newblock \emph{\bibinfo{journal}{Phys. Rep.}} \textbf{\bibinfo{volume}{461}},
  \bibinfo{pages}{111} (\bibinfo{year}{2008}).

\bibitem{LS99}
\bibinfo{author}{Lebowitz, J.~L.} \& \bibinfo{author}{Spohn, H.}
\newblock \bibinfo{title}{A {G}allavotti-{C}ohen-type symmetry in the large
  deviation functional for stochastic dynamics}.
\newblock \emph{\bibinfo{journal}{J. Stat. Phys.}}
  \textbf{\bibinfo{volume}{95}}, \bibinfo{pages}{333} (\bibinfo{year}{1999}).

\bibitem{seifert05}
\bibinfo{author}{Seifert, U.}
\newblock \bibinfo{title}{Entropy production along a stochastic trajectory and
  an integral fluctuation theorem}.
\newblock \emph{\bibinfo{journal}{Phys. Rev. Lett.}}
  \textbf{\bibinfo{volume}{95}}, \bibinfo{pages}{040602}
  (\bibinfo{year}{2005}).

\bibitem{kim2004entropy}
\bibinfo{author}{Kim, K.~H.} \& \bibinfo{author}{Qian, H.}
\newblock \bibinfo{title}{Entropy production of brownian macromolecules with
  inertia}.
\newblock \emph{\bibinfo{journal}{Physical review letters}}
  \textbf{\bibinfo{volume}{93}}, \bibinfo{pages}{120602}
  (\bibinfo{year}{2004}).

\bibitem{sarracino}
\bibinfo{author}{Sarracino, A.}
\newblock \bibinfo{title}{Time asymmetry of the kramers equation with nonlinear
  friction: Fluctuation-dissipation relation and ratchet effect}.
\newblock \emph{\bibinfo{journal}{Phys. Rev. E}} \textbf{\bibinfo{volume}{88}},
  \bibinfo{pages}{052124} (\bibinfo{year}{2013}).

\bibitem{ganguly2013stochastic}
\bibinfo{author}{Ganguly, C.} \& \bibinfo{author}{Chaudhuri, D.}
\newblock \bibinfo{title}{Stochastic thermodynamics of active brownian
  particles}.
\newblock \emph{\bibinfo{journal}{Physical Review E}}
  \textbf{\bibinfo{volume}{88}}, \bibinfo{pages}{032102}
  (\bibinfo{year}{2013}).

\bibitem{chaudhuri2014active}
\bibinfo{author}{Chaudhuri, D.}
\newblock \bibinfo{title}{Active brownian particles: Entropy production and
  fluctuation response}.
\newblock \emph{\bibinfo{journal}{Physical Review E}}
  \textbf{\bibinfo{volume}{90}}, \bibinfo{pages}{022131}
  (\bibinfo{year}{2014}).

\bibitem{FNCTVW16}
\bibinfo{author}{Fodor, E.} \emph{et~al.}
\newblock \bibinfo{title}{How far from equilibrium is active matter?}
\newblock \emph{\bibinfo{journal}{Phys. Rev. Lett.}}
  \textbf{\bibinfo{volume}{117}}, \bibinfo{pages}{038103}
  (\bibinfo{year}{2016}).

\bibitem{CP15}
\bibinfo{author}{Cerino, L.} \& \bibinfo{author}{Puglisi, A.}
\newblock \bibinfo{title}{Entropy production for velocity-dependent macroscopic
  forces: The problem of dissipation without fluctuations}.
\newblock \emph{\bibinfo{journal}{Europhys. Lett.}}
  \textbf{\bibinfo{volume}{111}}, \bibinfo{pages}{40012}
  (\bibinfo{year}{2015}).

\bibitem{BGJL13}
\bibinfo{author}{Bertini, L.}, \bibinfo{author}{Gabrielli, D.},
  \bibinfo{author}{Jona-Lasinio, G.} \& \bibinfo{author}{Landim, C.}
\newblock \bibinfo{title}{Clausius inequality and optimality of quasistatic
  transformations for nonequilibrium stationary states}.
\newblock \emph{\bibinfo{journal}{Phys. Rev. Lett.}}
  \textbf{\bibinfo{volume}{110}}, \bibinfo{pages}{020601}
  (\bibinfo{year}{2013}).

\bibitem{MN14}
\bibinfo{author}{Maes, C.} \& \bibinfo{author}{Neto\u{c}n\'y, K.}
\newblock \bibinfo{title}{A nonequilibrium extension of the clausius heat
  theorem}.
\newblock \emph{\bibinfo{journal}{J. Stat. Phys.}}
  \textbf{\bibinfo{volume}{154}}, \bibinfo{pages}{188} (\bibinfo{year}{2014}).

\bibitem{BDGJL15}
\bibinfo{author}{Bertini, L.}, \bibinfo{author}{Sole, A.~D.},
  \bibinfo{author}{Gabrielli, D.}, \bibinfo{author}{Jona-Lasinio, G.} \&
  \bibinfo{author}{Landim, C.}
\newblock \bibinfo{title}{Quantitative analysis of the clausius inequality}.
\newblock \emph{\bibinfo{journal}{J. Stat. Mech.}} \bibinfo{pages}{P10018}
  (\bibinfo{year}{2015}).

\bibitem{hanggi1995colored}
\bibinfo{author}{Hanggi, P.} \& \bibinfo{author}{Jung, P.}
\newblock \bibinfo{title}{Colored noise in dynamical systems}.
\newblock \emph{\bibinfo{journal}{Advances in Chemical Physics}}
  \textbf{\bibinfo{volume}{89}}, \bibinfo{pages}{239--326}
  (\bibinfo{year}{1995}).

\bibitem{jung1987dynamical}
\bibinfo{author}{Jung, P.} \& \bibinfo{author}{H{\"a}nggi, P.}
\newblock \bibinfo{title}{Dynamical systems: a unified colored-noise
  approximation}.
\newblock \emph{\bibinfo{journal}{Physical review A}}
  \textbf{\bibinfo{volume}{35}}, \bibinfo{pages}{4464} (\bibinfo{year}{1987}).

\bibitem{h1989colored}
\bibinfo{author}{H'walisz, L.}, \bibinfo{author}{Jung, P.},
  \bibinfo{author}{H{\"a}nggi, P.}, \bibinfo{author}{Talkner, P.} \&
  \bibinfo{author}{Schimansky-Geier, L.}
\newblock \bibinfo{title}{Colored noise driven systems with inertia}.
\newblock \emph{\bibinfo{journal}{Zeitschrift f{\"u}r Physik B Condensed
  Matter}} \textbf{\bibinfo{volume}{77}}, \bibinfo{pages}{471--483}
  (\bibinfo{year}{1989}).

\bibitem{PPRV10}
\bibinfo{author}{Puglisi, A.}, \bibinfo{author}{Pigolotti, S.},
  \bibinfo{author}{Rondoni, L.} \& \bibinfo{author}{Vulpiani, A.}
\newblock \bibinfo{title}{Entropy production and coarse-graining in markov
  processes}.
\newblock \emph{\bibinfo{journal}{J. Stat. Mech.}} \bibinfo{pages}{P05015}
  (\bibinfo{year}{2010}).

\bibitem{CPV12}
\bibinfo{author}{Crisanti, A.}, \bibinfo{author}{Puglisi, A.} \&
  \bibinfo{author}{Villamaina, D.}
\newblock \bibinfo{title}{Non-equilibrium and information: the role of
  cross-correlations}.
\newblock \emph{\bibinfo{journal}{Phys. Rev. E 85}}
  \textbf{\bibinfo{volume}{85}}, \bibinfo{pages}{061127}
  (\bibinfo{year}{2012}).

\bibitem{esposito2012stochastic}
\bibinfo{author}{Esposito, M.}
\newblock \bibinfo{title}{Stochastic thermodynamics under coarse graining}.
\newblock \emph{\bibinfo{journal}{Physical Review E}}
  \textbf{\bibinfo{volume}{85}}, \bibinfo{pages}{041125}
  (\bibinfo{year}{2012}).

\bibitem{celani2012anomalous}
\bibinfo{author}{Celani, A.}, \bibinfo{author}{Bo, S.},
  \bibinfo{author}{Eichhorn, R.} \& \bibinfo{author}{Aurell, E.}
\newblock \bibinfo{title}{Anomalous thermodynamics at the microscale}.
\newblock \emph{\bibinfo{journal}{Physical review letters}}
  \textbf{\bibinfo{volume}{109}}, \bibinfo{pages}{260603}
  (\bibinfo{year}{2012}).

\bibitem{bo2014entropy}
\bibinfo{author}{Bo, S.} \& \bibinfo{author}{Celani, A.}
\newblock \bibinfo{title}{Entropy production in stochastic systems with fast
  and slow time-scales}.
\newblock \emph{\bibinfo{journal}{Journal of Statistical Physics}}
  \textbf{\bibinfo{volume}{154}}, \bibinfo{pages}{1325--1351}
  (\bibinfo{year}{2014}).

\bibitem{maggi2015multidimensional}
\bibinfo{author}{Maggi, C.}, \bibinfo{author}{Marconi, U. M.~B.},
  \bibinfo{author}{Gnan, N.} \& \bibinfo{author}{Di~Leonardo, R.}
\newblock \bibinfo{title}{Multidimensional stationary probability distribution
  for interacting active particles}.
\newblock \emph{\bibinfo{journal}{Scientific Reports}}
  \textbf{\bibinfo{volume}{5}} (\bibinfo{year}{2015}).

\bibitem{risken}
\bibinfo{author}{Risken, H.}
\newblock \emph{\bibinfo{title}{Fokker-Planck Equation}}
  (\bibinfo{publisher}{Springer}, \bibinfo{year}{1984}).

\bibitem{PVBTW05}
\bibinfo{author}{Puglisi, A.}, \bibinfo{author}{Visco, P.},
  \bibinfo{author}{Barrat, A.}, \bibinfo{author}{Trizac, E.} \&
  \bibinfo{author}{van Wijland, F.}
\newblock \bibinfo{title}{Fluctuations of internal energy flow in a vibrated
  granular gas}.
\newblock \emph{\bibinfo{journal}{Phys. Rev. Lett.}}
  \textbf{\bibinfo{volume}{95}}, \bibinfo{pages}{110202}
  (\bibinfo{year}{2005}).

\bibitem{ge2014time}
\bibinfo{author}{Ge, H.}
\newblock \bibinfo{title}{Time reversibility and nonequilibrium thermodynamics
  of second-order stochastic processes}.
\newblock \emph{\bibinfo{journal}{Physical Review E}}
  \textbf{\bibinfo{volume}{89}}, \bibinfo{pages}{022127}
  (\bibinfo{year}{2014}).

\bibitem{bo2013entropic}
\bibinfo{author}{Bo, S.} \& \bibinfo{author}{Celani, A.}
\newblock \bibinfo{title}{Entropic anomaly and maximal efficiency of
  microscopic heat engines}.
\newblock \emph{\bibinfo{journal}{Physical Review E}}
  \textbf{\bibinfo{volume}{87}}, \bibinfo{pages}{050102}
  (\bibinfo{year}{2013}).

\bibitem{DEGM}
\bibinfo{author}{de~Groot, S.~R.} \& \bibinfo{author}{Mazur, P.}
\newblock \emph{\bibinfo{title}{Non-equilibrium thermodynamics}}
  (\bibinfo{publisher}{Dover Publications}, \bibinfo{address}{New York},
  \bibinfo{year}{1984}).

\bibitem{kreuzer1981nonequilibrium}
\bibinfo{author}{Kreuzer, H.~J.}
\newblock \bibinfo{title}{Nonequilibrium thermodynamics and its statistical
  foundations}.
\newblock \emph{\bibinfo{journal}{Oxford and New York, Clarendon Press, 1981.
  455 p.}} \textbf{\bibinfo{volume}{1}} (\bibinfo{year}{1981}).

\bibitem{marconi2016pressure}
\bibinfo{author}{Marconi, U. M.~B.}, \bibinfo{author}{Maggi, C.} \&
  \bibinfo{author}{Melchionna, S.}
\newblock \bibinfo{title}{Pressure and surface tension of an active simple
  liquid: a comparison between kinetic, mechanical and free-energy based
  approaches}.
\newblock \emph{\bibinfo{journal}{Soft Matter}}  (\bibinfo{year}{2016}).

\bibitem{fox1986functional}
\bibinfo{author}{Fox, R.~F.}
\newblock \bibinfo{title}{Functional-calculus approach to stochastic
  differential equations}.
\newblock \emph{\bibinfo{journal}{Physical Review A}}
  \textbf{\bibinfo{volume}{33}}, \bibinfo{pages}{467} (\bibinfo{year}{1986}).

\bibitem{chapman1970mathematical}
\bibinfo{author}{Chapman, S.} \& \bibinfo{author}{Cowling, T.~G.}
\newblock \emph{\bibinfo{title}{The mathematical theory of non-uniform gases:
  an account of the kinetic theory of viscosity, thermal conduction and
  diffusion in gases}} (\bibinfo{publisher}{Cambridge university press},
  \bibinfo{year}{1970}).

\bibitem{marconi2010dynamic}
\bibinfo{author}{Marconi, U. M.~B.} \& \bibinfo{author}{Melchionna, S.}
\newblock \bibinfo{title}{Dynamic density functional theory versus kinetic
  theory of simple fluids}.
\newblock \emph{\bibinfo{journal}{Journal of Physics: Condensed Matter}}
  \textbf{\bibinfo{volume}{22}}, \bibinfo{pages}{364110}
  (\bibinfo{year}{2010}).

\bibitem{fodor2016far}
\bibinfo{author}{Fodor, {\'E}.} \emph{et~al.}
\newblock \bibinfo{title}{How far from equilibrium is active matter?}
\newblock \emph{\bibinfo{journal}{arXiv preprint arXiv:1604.00953}}
  (\bibinfo{year}{2016}).

\bibitem{paoluzzi2016critical}
\bibinfo{author}{Paoluzzi, M.}, \bibinfo{author}{Maggi, C.},
  \bibinfo{author}{Marini Bettolo~Marconi, U.} \& \bibinfo{author}{Gnan, N.}
\newblock \bibinfo{title}{Critical phenomena in active matter}.
\newblock \emph{\bibinfo{journal}{Physical Review E}}
  \textbf{\bibinfo{volume}{94}}, \bibinfo{pages}{052602}
  (\bibinfo{year}{2016}).

\bibitem{marconi2006nonequilibrium}
\bibinfo{author}{Marconi, U. M.~B.} \& \bibinfo{author}{Tarazona, P.}
\newblock \bibinfo{title}{Nonequilibrium inertial dynamics of colloidal
  systems}.
\newblock \emph{\bibinfo{journal}{The Journal of Chemical Physics}}
  \textbf{\bibinfo{volume}{124}}, \bibinfo{pages}{164901}
  (\bibinfo{year}{2006}).

\bibitem{marini2007}
\bibinfo{author}{Marini Bettolo~Marconi, U.}, \bibinfo{author}{Tarazona, P.} \&
  \bibinfo{author}{Cecconi, F.}
\newblock \bibinfo{title}{Theory of thermostatted inhomogeneous granular
  fluids: A self-consistent density functional description}.
\newblock \emph{\bibinfo{journal}{The Journal of Chemical Physics}}
  \textbf{\bibinfo{volume}{126}}, \bibinfo{pages}{164904}
  (\bibinfo{year}{2007}).

\bibitem{titulaer1978systematic}
\bibinfo{author}{Titulaer, U.~M.}
\newblock \bibinfo{title}{A systematic solution procedure for the fokker-planck
  equation of a brownian particle in the high-friction case}.
\newblock \emph{\bibinfo{journal}{Physica A: Statistical Mechanics and its
  Applications}} \textbf{\bibinfo{volume}{91}}, \bibinfo{pages}{321--344}
  (\bibinfo{year}{1978}).

\bibitem{marconi2016effective}
\bibinfo{author}{Marconi, U. M.~B.}, \bibinfo{author}{Paoluzzi, M.} \&
  \bibinfo{author}{Maggi, C.}
\newblock \bibinfo{title}{Effective potential method for active particles}.
\newblock \emph{\bibinfo{journal}{Molecular Physics}} \bibinfo{pages}{1--11}
  (\bibinfo{year}{2016}).

\end{thebibliography}



\section*{ Contributions}

U.M.B.M., C.M. and A.P. contributed equally to the manuscript.

\section*{ Competing interests}

The authors declare no competing financial interests.

\end{document}